\documentclass[11pt,a4paper]{article}

\usepackage[T1]{fontenc}
\usepackage[utf8]{inputenc}
\usepackage{lmodern}
\usepackage{microtype}
\usepackage[a4paper,margin=2.35cm]{geometry}
\usepackage{amsmath,amssymb,mathtools,bm}
\usepackage{booktabs,tabularx,array,longtable}
\usepackage[dvipsnames]{xcolor}
\usepackage{enumitem}
\usepackage{tikz}
\usetikzlibrary{arrows.meta,positioning,fit,calc}
\usepackage[most]{tcolorbox}
\usepackage{hyperref}
\usepackage{cleveref}

\definecolor{deepblue}{RGB}{27,73,122}
\definecolor{teal}{RGB}{0,115,119}
\definecolor{warm}{RGB}{158,82,35}
\definecolor{pale}{RGB}{244,247,250}
\hypersetup{
  colorlinks=true,
  linkcolor=deepblue,
  citecolor=teal,
  urlcolor=warm,
  pdftitle={Critical emergence of quantum theory, spacetime and gravity from generalized trace dynamics},
  pdfauthor={Tejinder P. Singh}
}

\setlist{itemsep=0.25em,topsep=0.4em}
\newcolumntype{Y}{>{\raggedright\arraybackslash}X}
\newcolumntype{C}{>{\centering\arraybackslash}X}

\newcommand{\Tr}{\operatorname{Tr}}
\newcommand{\tr}{\operatorname{tr}}

\newcommand{\rank}{\operatorname{rank}}

\newcommand{\dd}{\mathrm d}
\newcommand{\ii}{\mathrm i}
\newcommand{\ee}{\mathrm e}

\newcommand{\cD}{\mathcal D}

\newcommand{\cN}{\mathcal N}
\newcommand{\cO}{\mathcal O}
\newcommand{\cP}{\mathcal P}
\newcommand{\cS}{\mathcal S}
\newcommand{\cV}{\mathcal V}
\newcommand{\eff}{\mathrm{eff}}
\newcommand{\TD}{\mathrm{TD}}
\newcommand{\loc}{\mathrm{loc}}
\newcommand{\crit}{\mathrm{crit}}
\newcommand{\Pl}{\mathrm{Pl}}
\newcommand{\AS}{\mathrm{AS}}
\newcommand{\SK}{\mathrm{SK}}

\newtcolorbox{statusbox}[1]{
  enhanced,breakable,colback=pale,colframe=deepblue,
  boxrule=0.7pt,arc=1.5mm,left=1.5mm,right=1.5mm,
  title={#1},fonttitle=\bfseries
}
\newtcolorbox{cautionbox}[1]{
  enhanced,breakable,colback=orange!4,colframe=warm,
  boxrule=0.7pt,arc=1.5mm,left=1.5mm,right=1.5mm,
  title={#1},fonttitle=\bfseries
}

\title{\textbf{Critical emergence of quantum theory, spacetime and gravity from generalized trace dynamics}\\[0.4em]
\large A quantitative condensed-matter framework for the $E_8\times\omega E_8$ programme}
\author{Tejinder P. Singh\\
\normalsize Tata Institute of Fundamental Research, Homi Bhabha Road, Mumbai 400005, India\\
\normalsize \href{mailto:tpsingh@tifr.res.in}{tpsingh@tifr.res.in}}
\date{9 August 2026}

\begin{document}
\maketitle

\begin{abstract}
Generalized trace dynamics (GTD) proposes a deterministic matrix dynamics of atoms of space-time-matter, or aikyons, whose near-equilibrium statistical mechanics yields quantum dynamics, while sufficiently large anti-self-adjoint fluctuations drive spontaneous localisation and the emergence of classical spacetime. Two logically independent large limits enter this proposal. Conditional quantum Ward identities arise from long-Connes-time coarse-graining of a finite trace-dynamical system---in principle even one aikyon with its matrix degrees of freedom---provided a stationary measure, mixing and the other trace-dynamical assumptions hold. A genuine collective singularity instead requires an extensive many-aikyon and/or large-matrix limit. Long observation time can estimate equilibrium averages but cannot by itself supply simultaneously interacting constituents. The $E_8\times\omega E_8$ programme further proposes that an electroweak-neighbourhood breaking reorganises the split-bioctonionic and $SO(3,3)$ scaffolding into four-dimensional gravitational and weak sectors. This paper turns the often-used analogy with condensed matter into a quantitative, falsifiable effective framework. We distinguish four coupled but logically separate channels: a bosonic bifermionic localisation composite, a geometric soldering or Lorentz--Higgs composite, a gauge-invariant electroweak composite, and a matrix-valued finite-internal or flavour bridge. A source-dependent many-aikyon partition function defines their exact susceptibility matrix and Legendre effective action. A single continuous common mode requires a nondegenerate zero eigenvalue of the zero-frequency inverse susceptibility with support in all required channels; a degenerate instability must instead be characterised by its basis-invariant critical spectral projector. A first-order common event requires a multi-channel jump together with phase free-energy equality. These criteria also show why quartic couplings inserted in a Landau potential do not by themselves establish a common transition.

We propose operational diagnostics for localisation, commutativity, simplicity, nonzero four-volume and spectral dimension; formulate the nonequilibrium problem in Schwinger--Keldysh language; define a projected fluctuation--dissipation defect; state the martingale condition required for Born probabilities; and identify Newton's constant with the long-wavelength stiffness of the ordered geometric phase. We connect the projected GTD kernel to objective-collapse benchmarks and to the dynamic structure factors measured in condensed matter, while keeping the collapse operator, amplification law and physical-time map as quantities to be derived. We distinguish Sakharov's induced action, the Wesley--Singh--Isidro constrained-$BF$ route to the Einstein--Hilbert action, Jacobson's local equation of state and Padmanabhan's entropy/holographic dynamics, and formulate the coefficient-matching conditions required for these descriptions to be limits of one GTD substrate. We compare this constitutive picture with Verlinde's de~Sitter elastic gravity and show how the programme's cosmological infrared action $S_{\rm IR}[g,T]$ realises a GR-to-MOND response crossover, while leaving its derivation from many-aikyon correlators open. We separate the trace-dynamical temperature, cosmological temperature and renormalisation scale, and derive the clock-matching relation needed to compare them. Continuous, first-order and crossover scenarios are treated separately, including Kibble--Zurek scaling and the conditional gravitational-wave scale of an electroweak-neighbourhood first-order event. We also explain why neither standard CSL dynamics nor the Standard Model electroweak crossover by itself supplies the required thermodynamic criticality. A recent finite-Dirac stress test conditionally disfavours, at stated inputs, one economical common Hermitian rank-two flavour truncation, motivating a full matrix order parameter. As a first execution of selected diagnostics, we study a regulated two-sector surrogate matrix ensemble: a Myers-deformed geometric triple coupled to a matter matrix, with exact global unitary invariance. In this surrogate the discretised trace-dynamics flow conserves the bosonic Adler--Millard charge to machine precision, stationary branch covariances and a transient multi-channel jump direction are measurable, hysteresis and branch discontinuities are consistent with first-order geometric ordering that drags the matter channel, the jump-direction estimator passes its built-in coupled-versus-decoupled null test, the classical equilibrium fluctuation--dissipation relation is verified and its quench violation quantified, and only the subsystem-projected charge carries broadband noise. A Grassmann-regulated extension then couples two exactly integrated degenerate matrix-fermion flavours to the dynamical geometry: the fermionic spectral-dimension flow identifies the emergent two-sphere ($d_s=1.96$ plateau) and refuses a dimension to the disordered phase, a fermion-derived mode-number channel joins the multi-channel jump, and promoting the determinant into the measure back-reacts strongly on the phase diagram while preserving charge to machine precision in the dedicated integrator check. These demonstrations establish that selected framework objects are calculable, not that the exact coexistence free energy has been located or that GTD itself passes the tests. The result is a calculational programme rather than a completed derivation. Its primary next step is unambiguous: construct the regulated many-aikyon measure and compute the connected composite two-point matrix and projected Adler--Millard fluctuation kernel for GTD itself.
\end{abstract}

\begin{statusbox}{Epistemic convention}
Throughout this paper, \emph{established} denotes a standard mathematical or physical result; \emph{programme input} denotes a stated ingredient of GTD or the $E_8\times\omega E_8$ construction; \emph{proposed} denotes a new effective definition or diagnostic introduced here; \emph{conditional} denotes a result following from displayed assumptions; and \emph{open} denotes a calculation not yet supplied. The status is part of the scientific content.
\end{statusbox}

\tableofcontents

\section{The question and the sharpened claim}
\label{sec:intro}

The condensed-matter analogy behind the present programme contains two distinct uses of statistical mechanics. Long-time self-averaging can supply a statistical description of a finite deterministic system, whereas an ordered phase requires an extensive set of simultaneously interacting degrees of freedom. In a superfluid, for example, equilibrium statistical mechanics and condensate hydrodynamics are not rival theories, but neither are the long observation-time limit and the thermodynamic limit the same operation. The analogy suggests the schematic pair
\begin{equation}
\boxed{
\begin{gathered}
\text{GTD near equilibrium}\longrightarrow\text{quantum dynamics},\\
\text{GTD localisation and ordering}\longrightarrow
\text{classical geometry and gravity}.
\end{gathered}
}
\label{eq:central_pair}
\end{equation}
Trace dynamics supplies the first arrow under canonical-ensemble and Ward-identity assumptions \cite{AdlerMillard1996,AdlerBook,Adler2023}. Earlier work in thermodynamic and induced-gravity approaches motivates the second arrow at a broad conceptual level \cite{Sakharov1968,Jacobson1995,Padmanabhan2010}. GTD attempts to obtain both from the same matrix substrate \cite{SinghUniverse2026}.

The first arrow can be posed for a finite aikyon system, including a single
aikyon with sufficiently rich matrix degrees of freedom, by coarse-graining
over many microscopic intervals of Connes time. The second arrow is
collective and requires many aikyons and/or a large-matrix limit. Symbolically,
\begin{equation}
\boxed{
\text{long-Connes-time limit at fixed }(N_{\rm aik},n)
\quad\neq\quad
\text{extensive limit }(N_{\rm aik},n)\to\infty .
}
\label{eq:two_limits_intro}
\end{equation}
The first limit can justify replacing a time average by a stationary ensemble
average if mixing holds. It does not manufacture the simultaneous
inter-aikyon correlations or extensivity required by the second.

The sharpened claim is deliberately narrower than the metaphor. The formation of a geometric phase may be far from equilibrium; the resulting ordered phase need not remain far from equilibrium. General relativity, if recovered, is the infrared constitutive dynamics of that ordered phase. Quantum fields then describe degrees of freedom that remain unlocalised, or the quantised fluctuations of collective modes, on the emergent background. Thus
\begin{equation}
\boxed{
\text{far-from-equilibrium selection of a phase}
\quad\neq\quad
\text{permanently far-from-equilibrium gravity}.
}
\label{eq:not_far_forever}
\end{equation}

Three events are often placed near the same electroweak scale in the programme:
\begin{enumerate}[label=(\roman*)]
\item departure from trace-dynamical equilibrium and spontaneous localisation;
\item selection of a non-degenerate four-dimensional geometric or $BF$ branch;
\item electroweak, triality and finite-internal symmetry breaking.
\end{enumerate}
Their coincidence is not yet derived. The purpose of this paper is to replace that coincidence by a well-posed calculation. We will construct a vector of composite operators, compute its connected susceptibility matrix in the many-aikyon ensemble, and ask whether one instability has support in all three channels.

Here $E_8\times\omega E_8$ denotes the split-complex exchange-graded pair used in the programme. It is not a tensor product of a group with a number and is not a new Lie group. Its role is kinematic and representation-theoretic until an invariant dynamical action on the physical fermion module is established \cite{Kaushik2026}.

\subsection{What this paper does not assume}

We do not assume that the Higgs expectation value is itself the order parameter for gravity. We do not equate the trace-dynamical Lagrange multiplier with a cosmological temperature. We do not identify an electroweak renormalisation scale with a thermodynamic temperature. We do not assume a continuous transition, and we do not assign a Hohenberg--Halperin universality class before spatial locality exists. Finally, we do not assume that localisation of matrices is sufficient to reconstruct a four-dimensional Lorentzian manifold.

These exclusions matter. Without them, a phenomenological Landau potential can always be written that reproduces a desired vacuum, but it would merely store the answer in its coefficients.

\section{A cross-disciplinary dictionary}
\label{sec:dictionary}

Table~\ref{tab:dictionary} gives the proposed translation. Several entries are analogies, not identities. In particular, the Planck stiffness is not asserted to be generated by an electroweak condensate. The microscopic GTD action already contains the Planck scale; the ordering calculation must show how that scale appears in the response of the selected phase.

\begin{table}[htbp]
\centering
\caption{Condensed-matter language and its proposed GTD counterpart. ``Diagnostic'' means an operational quantity defined in this paper, not a result already calculated from GTD.}
\label{tab:dictionary}
\small
\begin{tabularx}{\textwidth}{@{}p{0.27\textwidth}Yp{0.17\textwidth}@{}}
\toprule
Condensed-matter notion & GTD/$E_8\times\omega E_8$ counterpart & Standing\\
\midrule
Microscopic constituents & Aikyons or STM atoms, described by noncommuting bosonic and Grassmann-odd fermionic matrices evolving in Connes time & Programme input\\
Microscopic conserved charge & Adler--Millard charge from global unitary invariance & Established within trace dynamics\\
Normal phase & Regulated many-aikyon ensemble near canonical equilibrium & Conditional\\
Temporal coarse-graining & Long-Connes-time average of a finite trace-dynamical system, conditionally represented by a stationary ensemble under mixing/ergodicity & Conditional\\
Quantum statistical description & Ward identities and equipartition of the Adler--Millard charge; not intrinsically a many-aikyon phase transition & Conditional\\
Collective thermodynamic limit & $N_{\rm aik}\to\infty$ and/or matrix size $n\to\infty$, distinct from increasing the observation time & Required for a genuine singularity\\
Condensate amplitude & Even bifermionic/localisation composite and branch overlap & Proposed diagnostic\\
Order-parameter orientation & Leaf, real-form and Lorentz--Higgs/soldering selector & Programme input / open dynamics\\
Broken-phase frame & Co-frame obtained from a covariant Goldstone or constrained $BF$ variables & Conditional\\
Stiffness & Coefficient of the two-derivative geometric response, identified with $M_{\Pl}^2$ after matching & Proposed Kubo definition\\
Normal plus ordered components & Unlocalised quantum sector plus localised geometric/matter sector & Proposed two-component interpretation\\
Critical susceptibility & Connected matrix of localisation, geometry, Higgs and finite-internal composites & Proposed decisive calculation\\
Quench & Time-dependent passage of the control parameters through an instability & Conditional cosmological interpretation\\
Hydrodynamics & Diffeomorphism-invariant derivative expansion beginning with Einstein--Hilbert and matter terms & Conditional target\\
Material-specific corrections & Higher-curvature, nonlocal, stochastic and dissipative kernels & Open\\
\bottomrule
\end{tabularx}
\end{table}

\begin{figure}[htbp]
\centering
\begin{tikzpicture}[
node distance=9mm,
box/.style={draw=deepblue,rounded corners,fill=pale,text width=0.68\textwidth,align=center,inner sep=7pt},
arr/.style={-{Latex[length=2.3mm]},thick,deepblue}
]
\node[box] (l0) {\textbf{Level 0: microscopic matrix dynamics}\\Aikyons, $E_8\times\omega E_8$ labels, split-bioctonionic scaffold, Connes time};
\node[box,below=of l0] (l1) {\textbf{Level I: near-equilibrium statistics}\\Conditional Adler--Millard Ward identities and emergent quantum dynamics};
\node[box,below=of l1] (l3) {\textbf{Level III: ordered classical phase}\\Localised matter, non-degenerate co-frame, four-dimensional geometry, gravitational hydrodynamics};
\node[box,below=of l3] (l2) {\textbf{Level II: useful hybrid approximation}\\Quantum fields from Level I on the classical background from Level III};
\draw[arr] (l0) -- node[right,align=left]{long-Connes-time coarse-graining\\at fixed finite content} (l1);
\draw[arr] (l1) -- node[right,align=left]{collective instability in the\\large-aikyon/matrix limit} (l3);
\draw[arr] (l1.east) to[out=0,in=0,looseness=1.15] node[right,align=left]{unlocalised\\sector} (l2.east);
\draw[arr] (l3) -- node[right]{background limit} (l2);
\end{tikzpicture}
\caption{Four levels of description. The first arrow is temporal
self-averaging and may be posed for a finite aikyon system; the second requires
an extensive collective limit. Neither arrow is a change of spatial location.
Before the geometric transition there is no assumed physical space in which an
ordinary correlation length can be measured.}
\label{fig:levels}
\end{figure}
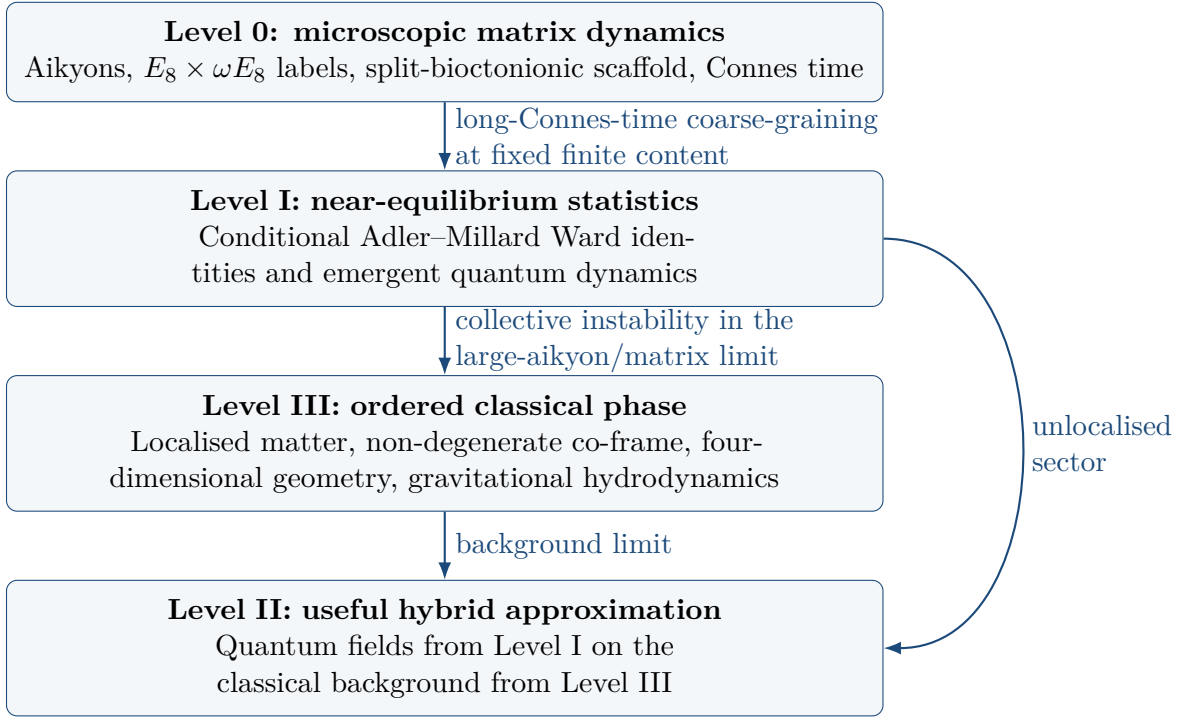

The proposed hierarchy is shown in Fig.~\ref{fig:levels}. The intermediate description, quantum field theory on classical curved spacetime, is a hybrid: quantum matter is inherited from the near-equilibrium sector, while the background is inherited from the ordered sector.

\section{Microscopic statistical starting point}
\label{sec:micro}

\subsection{Trace dynamics in one page}

Let $q_r(\tau)$ be bosonic or fermionic matrix variables and let
\begin{equation}
L_{\Tr}=\Tr\,L(q_r,\dot q_r),
\qquad
p_r=\frac{\delta L_{\Tr}}{\delta \dot q_r}.
\end{equation}
Cyclicity under the trace permits operator Euler--Lagrange equations. If the trace Hamiltonian is invariant under global unitary conjugation, Noether's theorem gives the Adler--Millard charge
\begin{equation}
\widetilde C
=\sum_{r\in B}[q_r,p_r]-\sum_{r\in F}\{q_r,p_r\}.
\label{eq:AM}
\end{equation}
The canonical ensemble has the schematic form
\begin{equation}
\rho_{\TD}=Z^{-1}
\exp\!\left[-\beta_{\TD}H_{\Tr}
-\Tr(\Lambda\widetilde C)-\mu_F N_F\right],
\label{eq:ensemble}
\end{equation}
where the precise set of conserved constraints depends on the regulated model. In ordinary trace dynamics, equipartition and Ward identities can yield effective canonical commutators and unitary quantum evolution, provided the measure exists, the relevant surface terms vanish, a scale hierarchy suppresses trace-Hamiltonian variations, and bosonic/fermionic balance is adequate \cite{AdlerMillard1996,AdlerBook,Adler2023}.

In GTD the degrees of freedom include space-time-matter matrices and evolve with respect to Connes time \cite{LochanSingh2011,LochanSatinSingh2012,ConnesRovelli1994}. The specific $E_8\times\omega E_8$ programme uses dotted and undotted bosonic and fermionic variables with left/right components. The bosonic zeroth quaternionic modes are separated from the differential Dirac part and seed scalar channels, while the finite/internal fermionic Dirac block requires no analogous subtraction \cite{SinghUniverse2026}. These bookkeeping details constrain the composites used below.

\begin{cautionbox}{What equilibrium does and does not do}
The ensemble does not create Planck's constant from nothing. In the current GTD formulation, $\hbar$ is a fundamental action scale and the equilibrium construction transmits it to effective commutators. Nor does the mere existence of $\widetilde C$ prove convergence of $Z$, ergodicity, locality, a positive Hilbert-space representation, or the emergence of quantum field theory. Those are separate acceptance tests.
\end{cautionbox}

\subsection{Two independent large limits: many times are not many aikyons}
\label{sec:two_limits}

The statistical step that yields effective quantum dynamics and the
thermodynamic step that could yield a critical geometric phase must be kept
separate. Let $\tau_{\rm corr}$ be a finite Connes-time correlation or mixing
scale of a regulated trace-dynamical system away from critical slowing. For an
observable $\mathcal A$ define
\begin{equation}
\overline{\mathcal A}_{\Delta\tau}
=\frac{1}{\Delta\tau}
\int_{\tau_0}^{\tau_0+\Delta\tau}\dd\tau\,\mathcal A(\tau),
\qquad
\Delta\tau=N_\tau\tau_{\rm corr},\qquad N_\tau\gg1 .
\label{eq:connes_time_average}
\end{equation}
Here $N_\tau$ counts correlation times; it is neither the number of aikyons nor
the matrix size. We use $\tau_{\rm corr}$ rather than the notation
$\tau_{\rm crit}$ to avoid confusing this microscopic sampling scale with a
thermodynamic critical time.
If the microscopic flow possesses the stationary measure
Eq.~\eqref{eq:ensemble} and is sufficiently mixing on the sector being
sampled, then
\begin{equation}
\lim_{\Delta\tau/\tau_{\rm corr}\to\infty}
\overline{\mathcal A}_{\Delta\tau}
=\langle\mathcal A\rangle_{\rho_{\TD}} .
\label{eq:ergodic_replacement}
\end{equation}
This is a temporal self-averaging statement. It can in principle be formulated
for a finite system, even a single aikyon with its internal matrix degrees of
freedom. It is the relevant large number behind the conditional
trace-dynamical Ward-identity derivation of effective quantum commutators.
Equation~\eqref{eq:ergodic_replacement} is an assumption to be tested, not an
automatic consequence of deterministic evolution: conserved sectors,
metastability or nonergodicity can invalidate it. A single aikyon must also
possess enough effective internal degrees of freedom to support mixing and the
canonical description; long duration alone does not ensure
self-thermalisation, and Eqs.~\eqref{eq:connes_time_average} and
\eqref{eq:ergodic_replacement} do not replace the other Ward-identity
assumptions stated above.

A thermodynamic transition asks a different question. At fixed aikyon number
$N_{\rm aik}$ and finite matrix size $n$, a regulated partition function is
normally analytic and arbitrarily long observation only estimates its finite
correlators more accurately. A nonanalytic transition or spontaneous order
requires an extensive limit,
\begin{equation}
(N_{\rm aik},n)\longrightarrow\infty
\quad\hbox{in a specified scaling and regulator order}.
\label{eq:extensive_limit}
\end{equation}
Thus
\begin{equation}
\mathcal L_{\tau}
:=\lim_{\Delta\tau/\tau_{\rm corr}\to\infty}
\Big|_{N_{\rm aik},n\ {\rm fixed}},
\qquad
\mathcal L_{\rm ext}
:=\lim_{N_{\rm aik},\,n\to\infty},
\qquad
\mathcal L_{\tau}\not\equiv\mathcal L_{\rm ext}.
\label{eq:limits_not_equal}
\end{equation}
Successive time blocks are correlated configurations of the same dynamical
system, not additional simultaneously interacting constituents. They may
replace ensemble \emph{sampling} under Eq.~\eqref{eq:ergodic_replacement};
they cannot replace the degrees of freedom whose correlations generate an
extensive instability. One could evade this conclusion only by constructing a
different theory in which history blocks themselves become coupled physical
variables. No such history-space construction is assumed here.

There is a logically possible third route: an infinite-time dynamical
large-deviation transition of time-integrated observables. Such a transition
would be defined by a tilted history ensemble and a nonanalytic scaled
cumulant-generating function, not by silently reinterpreting the static
$Z[J]$, $\Gamma[\Phi]$ and $\chi_{AB}(0)$ used here. Constructing that
history-space thermodynamics could be interesting, especially if one seeks a
literal many-times analogue of many molecules, but it is outside the present
framework and cannot be claimed without new variables and an explicit history
measure.

Operationally, the limits must be controlled in stages. At each finite
$(N_{\rm aik},n)$ one measures stationary correlators for
$\Delta\tau\gg\tau_{\rm corr}(N_{\rm aik},n;g)$, then performs finite-size
scaling toward Eq.~\eqref{eq:extensive_limit}, and removes a
symmetry-selecting source only after the extensive limit. Near a continuous
transition $\tau_{\rm corr}$ itself grows through critical slowing down, so
the temporal and extensive limits cannot be interchanged without analysis.
The water analogy maps most directly onto Eq.~\eqref{eq:extensive_limit}:
many molecules produce fluid or crystalline collective order. The
long-Connes-time averaging that produces the quantum Ward identities is a
separate, trace-dynamical use of statistical mechanics.

\subsection{The exact source construction}

Choose a set of bosonic composite operators $\cO_A(\tau)$ and introduce sources $J_A(\tau)$. The regulated generating functional is
\begin{align}
Z[J]&=\int \dd\mu_{\rm reg}[q,p]\,
\exp\!\left[-\beta_{\TD}H_{\Tr}-\Tr(\Lambda\widetilde C)-\mu_FN_F
+\int\dd\tau\,J_A\cO_A\right],\\
W[J]&=\ln Z[J],
\qquad
\Phi_A(\tau)=\frac{\delta W}{\delta J_A(\tau)},\\
\Gamma[\Phi]&=\sup_J\left\{\int\dd\tau\,J_A\Phi_A-W[J]\right\}.
\label{eq:Legendre}
\end{align}
At a translationally invariant equilibrium in Connes time, the connected susceptibility and the Hessian are
\begin{equation}
\chi_{AB}(\omega)
=\int\dd\tau\,\ee^{\ii\omega\tau}
\langle\delta\cO_A(\tau)\delta\cO_B(0)\rangle_c,
\qquad
\Gamma^{(2)}_{AB}(\omega)=\chi^{-1}_{AB}(\omega),
\label{eq:chiGamma}
\end{equation}
when the inverse exists on the physical, gauge-fixed or invariant subspace. Equation~\eqref{eq:chiGamma}, not a guessed polynomial potential, is the central quantitative bridge to condensed-matter language.

These convexity statements presume a real positive regulated measure and real sources for Hermitian scalar composites. If the microscopic construction instead requires a complex contour or has a sign problem, $\chi$ need not define a positive operator metric and the instability criterion must be reformulated in terms of physical retarded poles or a positive reconstructed measure. For a positive measure, $W[J]$ and its exact Legendre--Fenchel transform $\Gamma[\Phi]$ are convex. This fact is important at a first-order transition: the exact thermodynamic $\Gamma$ develops a Maxwell-flat coexistence face rather than two separated local minima with a barrier. Metastability is described by phase-restricted free energies, a constrained large-deviation functional or a coarse-grained effective action before convexification.

Before spacetime forms, Eq.~\eqref{eq:Legendre} is a many-matrix statistical problem in Connes time. A local functional $\int\dd^4x\,\mathcal L$ is not yet available. Matrix-index, graph or spectral notions of range may exist, but they must not be silently identified with physical distance.

\section{The required composite order parameters}
\label{sec:orders}

No single scalar currently captures the proposed transition. The minimal honest description contains at least four channels,
\begin{equation}
\bm\Phi=\left(\Phi_{\loc},\,\Phi_g,\,\Phi_H,\,\Phi_F\right),
\label{eq:vector_order}
\end{equation}
representing localisation, geometry, the electroweak composite and the finite-internal/flavour bridge. They may ultimately be projections of one master matrix composite. Treating them separately at first avoids building the desired coincidence into the notation.

\subsection{Localisation: an even composite and a branch overlap}

The fundamental fermionic variables are Grassmann odd. An ordinary c-number expectation value $\langle q_F\rangle$ is therefore not the appropriate condensate. A candidate order parameter must be Grassmann even. Schematically,
\begin{equation}
\Phi_{\loc,IJ}
=\left\langle
\cP_{\rm even}\!\left(q_{F,I}\,\mathcal K\,q_{F,J}\right)
\right\rangle,
\label{eq:bilinear}
\end{equation}
where $\mathcal K$ includes the regulator, adjoint and fixed auxiliary-Grassmann insertions required by the GTD action. The exact projection is open and must be derived rather than chosen for convenience. The bifermionic sector identified in the spectral-action decomposition supplies a natural operator family \cite{SinghUniverse2026}.

Localisation is branch selection. If the full ensemble averages over symmetry-related outcomes, a one-point function can vanish even when every ergodic component is localised. The situation resembles a ferromagnet averaged over domains or a spin glass averaged over pure states. Let $\alpha$ label an ergodic or outcome sector and $Q_A$ denote the bosonic spectral variables that become classical. Define the branch-sensitive overlap
\begin{equation}
q_{\loc}
=\frac{1}{N}\sum_{A=1}^{N}
\frac{\Tr\!\left(\langle Q_A\rangle_{\alpha}^{\dagger}
\langle Q_A\rangle_{\alpha}\right)}
{\left\langle\Tr(Q_A^{\dagger}Q_A)\right\rangle_{\alpha}}.
\label{eq:qloc}
\end{equation}
Equivalently, one may use a two-replica overlap with the replicas constrained to the same ergodic component. The value $q_{\loc}>0$ diagnoses branch freezing; it does not by itself prove that the frozen matrices define geometry.

There are two technical qualifications. First, if global unitary conjugation is a redundancy, two replicas must be aligned by extremising their relative-unitary overlap, or replaced by invariant spectral data; otherwise a small overlap can merely reflect an unremoved group orbit. Second, branch freezing is defined by an order of limits: a symmetry-selecting source is removed only after the many-aikyon and matrix-size limits. A nonzero value in one finite simulation is therefore a diagnostic, not proof of spontaneous ordering.

A complementary diagnostic is normalised noncommutativity,
\begin{equation}
\cN_{\rm nc}
=\frac{\sum_{A<B}\left\langle
\Tr[Q_A,Q_B]^{\dagger}[Q_A,Q_B]\right\rangle}
{\left(\sum_A\langle\Tr Q_A^{\dagger}Q_A\rangle\right)^2}.
\label{eq:Nnc}
\end{equation}
Classicalisation requires $\cN_{\rm nc}\to0$ in the selected sector, but commuting matrices alone may describe a set of eigenvalues with no four-dimensional locality.

\subsection{Geometry: soldering, simplicity and non-degeneracy}

The $SO(3,3)$ $BF$ construction proposes a symmetry-breaking field or normal that selects chiral four-dimensional sectors \cite{WesleySinghIsidro2026}. It combines the ideas of spontaneous soldering, constrained $BF$ gravity and graviweak symmetry breaking \cite{Percacci1984,FreidelStarodubtsev2005,Krasnov2011,NestiPercacci2008}. Write the geometric collective field as
\begin{equation}
\Phi_g=\rho_g U,
\label{eq:geomOP}
\end{equation}
where $U$ is a coset or branch orientation and $\rho_g$ is an invariant amplitude. In a compensator description, the co-frame is the covariant Goldstone field
\begin{equation}
e^a=\cP_{G/H}(DU)^a.
\label{eq:coframe}
\end{equation}
The exact group, real form and projection depend on the selected $SO(3,3)$ reduction. Equation~\eqref{eq:coframe} is therefore a target relation, not a derivation of soldering.

Two logically distinct geometric constructions must not be conflated. In the compensator route, $U$ is a zero-form whose covariant derivative supplies a candidate co-frame. In the Plebanski route, $B^i$ is a two-form constrained to be simple, locally $B^i\sim P^i{}_{ab}e^a\wedge e^b$. The programme must either derive this relation from one master composite or choose one route as fundamental. Simultaneously writing $e=DU$ and imposing simplicity on an unrelated $B$ field would only place the desired geometry in the ansatz.

Four separate tests are required:
\begin{align}
\text{simplicity:}\quad
\cS^{ij}&=B^i\wedge B^j-\frac{1}{3}\delta^{ij}B^k\wedge B_k\longrightarrow0,
\label{eq:simplicity}\\
\text{non-degeneracy:}\quad
\cV_4&=\frac{1}{4!}\epsilon_{abcd}\,
e^a\wedge e^b\wedge e^c\wedge e^d\neq0,
\label{eq:volume}\\
\text{reality:}\quad
g_{\mu\nu}&=e^a_{\mu}e^b_{\nu}\eta_{ab}\ \text{is real and Lorentzian},
\label{eq:reality}\\
\text{dimension:}\quad
N_D(\Lambda)&=\#\{|\lambda_n(D)|<\Lambda\}\sim C\Lambda^{d_s},
\qquad
d_s=\frac{\dd\ln N_D}{\dd\ln\Lambda}\longrightarrow4.
\label{eq:weyl}
\end{align}
The last test is to be applied to a regulated Euclidean continuation or another elliptic spectral problem; a Lorentzian Dirac operator does not directly have the required discrete Weyl counting function. The use of Dirac spectral data as geometric observables has an established precedent \cite{LandiRovelli1997}.
For a numerical calculation the form-valued simplicity condition should be converted into a dimensionless residual, for example
\begin{equation}
\mathfrak s_B=
\frac{\int\!\left\langle\cS^{ij}\wedge\star\cS_{ij}\right\rangle}
{\int\!\left\langle(B^k\wedge B_k)\wedge\star(B^\ell\wedge B_\ell)\right\rangle+\varepsilon_B}
\longrightarrow0,
\label{eq:simplicity_residual}
\end{equation}
where $\varepsilon_B$ regulates a vanishing denominator. The Hodge operator here can be used only after a provisional nondegenerate branch supplies a metric, so the test is necessarily iterative or must be recast algebraically before geometry emerges.
The set
\begin{equation}
q_{\loc}>0,\qquad \cN_{\rm nc}\to0,\qquad
\cS^{ij}\to0,\qquad \langle|\cV_4|\rangle>0,\qquad d_s\to4
\label{eq:geom_acceptance}
\end{equation}
is a substantially stronger acceptance criterion than ``the matrices have localised.'' It still leaves locality, topology, differentiability and the Einstein constitutive law to be demonstrated.

\subsection{Electroweak ordering: use gauge-invariant composites}

The local Higgs field $H$ is gauge variant, so an unfixed-gauge one-point expectation is not an observable order parameter. Gauge-invariant diagnostics include
\begin{equation}
\Phi_H=\langle H^{\dagger}H\rangle,
\qquad
\chi_H=\int\dd^4x\,
\langle(H^{\dagger}H)(x)(H^{\dagger}H)(0)\rangle_c,
\label{eq:HiggsInv}
\end{equation}
and suitably dressed correlators. This is the appropriate language in light of Elitzur's theorem \cite{Elitzur1975}.

Even $H^\dagger H$ is a diagnostic, not automatically a strict thermodynamic order parameter. Gauge--Higgs theories can possess analytically connected confinement-like and Higgs-like regions with no symmetry-breaking boundary \cite{FradkinShenker1979}. A singular susceptibility, coexistence discontinuity or independently identified global/discrete symmetry is required before the enlarged-sector event is called a phase transition.

With the observed Higgs mass, the Standard Model thermal electroweak change is a crossover, not a true phase transition; lattice calculations find a sharply located crossover near $159.5\pm1.5$ GeV \cite{Donofrio2014}. A genuine critical or first-order event in the present programme must therefore lie in the enlarged bifermionic, two-Higgs, triality, geometric or mirror sector. The Standard Model crossover can be its low-energy shadow, but cannot be cited as evidence that the enlarged transition exists.

\subsection{Finite-internal and flavour ordering}

The finite Dirac operator has a grading-odd structure
\begin{equation}
D_F^{(f)}=
\begin{pmatrix}0&Y_f\\Y_f^{\dagger}&0\end{pmatrix}
\oplus J_F
\begin{pmatrix}0&Y_f\\Y_f^{\dagger}&0\end{pmatrix}J_F^{-1}.
\label{eq:DF}
\end{equation}
Jordan spectral data can constrain singular values or root-mass moduli, but do not by themselves select the left and right frames of $Y_f$. The current CKM analysis makes this separation explicit: family-vacuum selection, chiral projection and right-frame locking remain open \cite{SinghCKM2026}. The associated order parameter $\Phi_F$ must therefore be matrix valued at the microscopic stage. Reducing it to a single triplet, one phase or a common rank-two plane would be an additional dynamical result, not a legitimate starting assumption.

\section{The decisive common-transition test}
\label{sec:hessian}

\subsection{Continuous instability}

Let $g$ denote a control parameter of the regulated ensemble. It could be $\beta_{\TD}$, an entanglement density, a matrix coupling, aikyon number density, or a combination. A continuous transition is defined only in a thermodynamic or large-matrix limit. Its zero-mode criterion is
\begin{equation}
\boxed{
\lim_{N_{\rm aik},\,n\to\infty}
\lambda_{\min}\!\left[\Gamma^{(2)}_{AB}(\omega=0;g_c)\right]=0,
}
\label{eq:zero_mode}
\end{equation}
with the regulator removed in a stated order. At finite aikyon number and matrix size, an exact zero is neither expected nor required. Let $P_{\crit}$ denote the spectral projector onto the eigenvalues that vanish in this limit. If $\rank P_{\crit}=1$, write $P_{\crit}=|v_{\crit}\rangle\langle v_{\crit}|$ in the chosen operator metric. A single mixed localisation--geometry--electroweak mode is then established only if
\begin{equation}
\|P_{\loc}v_{\crit}\|\,
\|P_gv_{\crit}\|\,
\|P_Hv_{\crit}\|>0,
\label{eq:support_condition}
\end{equation}
and, if finite-internal selection is claimed to occur at the same event, also $\|P_Fv_{\crit}\|>0$. If $\rank P_{\crit}>1$, an individual eigenvector is basis dependent and Eq.~\eqref{eq:support_condition} must be replaced by basis-invariant weights such as $\Tr(P_A P_{\crit})$. A degenerate critical subspace can describe several independently soft sectors at the same coupling; it is not evidence for one mixed mode unless microscopic symmetry or relations also fix their mixing.

Because rescaling a composite rescales Hessian components, channel support is meaningful only after fixing the operator normalisation or a positive operator-space metric $G_{AB}$, for example by unit ultraviolet covariance. The spectral and channel projectors and all norms are then defined with $G_{AB}$.

The susceptibility form is often more practical. The transition is signalled by one or more divergent eigenvalues of $\chi_{AB}(0)$, while the associated spectral projector identifies the soft subspace. In a finite regulated matrix model, one looks for finite-size scaling of the susceptibility peak and Binder-type cumulants, then studies the large-matrix and many-aikyon limits.

The limit in Eq.~\eqref{eq:zero_mode} is the extensive limit of
Sec.~\ref{sec:two_limits}, not the long observation-time limit. The latter is
needed to estimate each finite-size susceptibility reliably. Increasing
$\Delta\tau$ at fixed $(N_{\rm aik},n)$ cannot turn a smooth finite-system
response into a thermodynamic singularity.

\subsection{A symmetry obstruction to a naive common mode}

Suppose the symmetric phase has exact group $G$ and $\cO_A$, $\cO_B$ lie in inequivalent irreducible representations. Then symmetry gives
\begin{equation}
\chi_{AB}(0)=0
\label{eq:selection_chi}
\end{equation}
unless their tensor product contains the singlet with the required quantum numbers. Consequently, a single mixed quadratic zero mode cannot be assumed across arbitrary localisation, geometric and Higgs composites.

There are only three clean ways around this obstruction:
\begin{enumerate}
\item a \emph{master composite} $\mathbb M$ transforms irreducibly in the microscopic theory, while $\Phi_{\loc}$, $\Phi_g$ and $\Phi_H$ are projections that become distinct only after symmetry breaking;
\item an exact microscopic relation or enlarged symmetry ties the susceptibilities and enforces a multicritical degeneracy;
\item the transition is first order, so several order parameters jump together without a common Gaussian zero mode.
\end{enumerate}
Accidental equality of separate critical couplings is possible but is a tuning, not an explanation.

\subsection{What a Landau potential can and cannot show}

After choosing invariant amplitudes, a phenomenological potential may be written as
\begin{align}
V_{\eff}={}&r_{\ell}\tr L^{\dagger}L
+u_{\ell,1}(\tr L^{\dagger}L)^2
+u_{\ell,2}\tr(L^{\dagger}LL^{\dagger}L)
+r_g\rho_g^2+u_g\rho_g^4\nonumber\\
&+r_H H^{\dagger}H+u_H(H^{\dagger}H)^2
+\kappa_{\ell g}\tr(L^{\dagger}L)\rho_g^2
+\kappa_{\ell H}\tr(L^{\dagger}L)H^{\dagger}H
+\kappa_{gH}\rho_g^2H^{\dagger}H+\cdots,
\label{eq:Landau}
\end{align}
where $L$ is an even localisation matrix. This potential is useful for classifying phases once its coefficients are computed or constrained. It does not prove a common transition. At the fully symmetric origin, the quartic cross-couplings do not generate off-diagonal entries in the Hessian. If $L$ condenses first, however, it shifts the secondary masses,
\begin{equation}
r_g^{\eff}=r_g+\kappa_{\ell g}\langle\tr L^{\dagger}L\rangle,
\qquad
r_H^{\eff}=r_H+\kappa_{\ell H}\langle\tr L^{\dagger}L\rangle.
\label{eq:induced_masses}
\end{equation}
This can trigger subsequent transitions or, with sufficiently strong couplings, one first-order jump. The distinction is empirical and calculational.

For stability, the quartic form must be copositive on the physical invariant cone; taking all self-couplings positive is not sufficient if cross-couplings are negative. Gauge, reality and Grassmann-regulator constraints further restrict the allowed domain.

For a first-order common event, introduce phase-restricted free-energy densities $f_1(g)$ and $f_2(g)$, or equivalently a constrained large-deviation functional $I_g(\bm\Phi)$ before convexification. The coexistence test is
\begin{equation}
f_1(g_c)=f_2(g_c),
\qquad
\bm\Phi_1\neq\bm\Phi_2,
\qquad
I_{g_c}(\bm\Phi_1)=I_{g_c}(\bm\Phi_2)=\min I_{g_c},
\label{eq:first_order_coexistence}
\end{equation}
where either the free-energy equality or the equivalent large-deviation equality is sufficient; both are shown to make the relation explicit. The exact convex Legendre effective action has a flat face joining $\bm\Phi_1$ and $\bm\Phi_2$ at coexistence, not two positive-Hessian minima. The discontinuity $\Delta\bm\Phi=\bm\Phi_2-\bm\Phi_1$ must have nonzero invariant projections in every channel claimed to change. At finite size, bimodality, Binder behaviour and hysteresis are evidence for first-order dynamics, but the exact criterion additionally requires a controlled thermodynamic limit and a free-energy or equal-weight determination.

\section{Nonequilibrium dynamics and the origin of localisation}
\label{sec:noneq}

\subsection{Schwinger--Keldysh effective action}

An equilibrium free energy cannot describe noise, dissipation, causality and entropy production by itself. Double the collective fields on a closed time contour and define $r/a$ combinations. At quadratic order,
\begin{equation}
\Gamma_{\SK}^{(2)}
=\int\frac{\dd\omega}{2\pi}
\left[
\Phi_a^A(-\omega)K^R_{AB}(\omega)\Phi_r^B(\omega)
+\frac{\ii}{2}\Phi_a^A(-\omega)N_{AB}(\omega)\Phi_a^B(\omega)
\right].
\label{eq:SK}
\end{equation}
Here $K^A=(K^R)^\dagger$ follows from the reality condition and is not a second independent cross term; equivalently one may write the retarded and advanced terms symmetrically with an overall factor $1/2$. Causality, reality and positivity constrain $K^{R,A}$ and $N$. In local thermal equilibrium, dynamical KMS symmetry enforces fluctuation--dissipation and Onsager relations \cite{Crossley2017,Glorioso2017}. Before spacetime emerges, Eq.~\eqref{eq:SK} is a $0+1$ dimensional collective matrix functional in Connes time. A spacetime Schwinger--Keldysh EFT becomes legitimate only after the locality and soldering tests in Eq.~\eqref{eq:geom_acceptance} pass.

\subsection{Projected Adler--Millard fluctuation and an FDT defect}

Let $C_{\AS}$ be the subsystem-projected anti-self-adjoint fluctuation channel of the Adler--Millard charge, after all regulator and coarse-graining projections have been specified. The Hermitian matrix fluctuation is $-\ii\,\delta C_{\AS}$. Choose a fixed Hermitian test matrix $u$, or another invariant scalar contraction, and define the scalar coordinate
\begin{equation}
X_{\AS,u}=\Tr\!\left[u(-\ii\,\delta C_{\AS})\right].
\label{eq:Xcontraction}
\end{equation}
The label $u$ will be suppressed below. A matrix-valued treatment instead requires component indices and a specified positive matrix norm. Define
\begin{align}
F_{\AS}(\omega)
&=\frac12\int\dd\tau\,\ee^{\ii\omega\tau}
\left\langle\{X_{\AS}(\tau),X_{\AS}(0)\}\right\rangle,
\label{eq:Fsym}\\
G^R_{\AS}(\omega)
&=-\ii\int_0^{\infty}\dd\tau\,\ee^{\ii\omega\tau}
\left\langle[X_{\AS}(\tau),X_{\AS}(0)]\right\rangle,
\qquad
\rho_{\AS}=-2\,\mathrm{Im}\,G^R_{\AS}.
\label{eq:GR}
\end{align}
With these sign conventions, equilibrium KMS implies
\begin{equation}
F_{\AS}(\omega)
=\frac12\coth\!\left(\frac{\beta_{\TD}\hbar\omega}{2}\right)\rho_{\AS}(\omega).
\label{eq:FDT}
\end{equation}
A qualification is essential. Equation~\eqref{eq:FDT} is the quantum KMS relation for the effective quantum sector obtained only after the trace-dynamical Ward-identity and scale-hierarchy assumptions have been imposed. The underlying canonical ensemble of matrix phase-space variables obeys the corresponding classical fluctuation--response relation, recovered as $\beta_{\TD}\hbar\omega\ll1$. A surrogate Hamiltonian calculation therefore tests this classical limit, not by itself the full quantum $\coth$ relation.
A dimensionless diagnostic of disequilibrium is
\begin{equation}
\cD_{\rm FDT}
=\frac{\int\dd\omega\,w(\omega)
\left|F_{\AS}(\omega)-\tfrac12\coth(\beta_{\TD}\hbar\omega/2)\rho_{\AS}(\omega)\right|^2}
{\int\dd\omega\,w(\omega)|F_{\AS}(\omega)|^2},
\label{eq:DFDT}
\end{equation}
where $w(\omega)$ specifies the experimentally or dynamically relevant band. Then $\cD_{\rm FDT}=0$ at the assumed KMS fixed point. A nonzero value quantifies departure from equilibrium; it does not by itself produce a norm-preserving collapse law.

The projection in this definition is essential. The total Adler--Millard charge is conserved, so its unprojected global correlator has only a zero-frequency conservation contribution and cannot by itself furnish broadband collapse noise. One must specify a subsystem--environment split, derive the exchange fluctuation $\delta C_{\AS}^{\rm sub}$ and retain the associated memory and response kernels. Treating the conserved total charge as an externally prescribed white-noise source would be inconsistent.

\subsection{Born probabilities are a martingale condition}

For a two-branch state let $p(\tau)$ be the norm weight of one branch. The most general one-dimensional It\^o diffusion is
\begin{equation}
\dd p=a(p,\tau)\dd\tau+b(p,\tau)\dd W_{\tau}.
\label{eq:dp}
\end{equation}
If $p$ reaches $0$ or $1$ with probability one and
\begin{equation}
a(p,\tau)=0,
\label{eq:martingale}
\end{equation}
then $p$ is a bounded martingale and the probability of absorption at $1$ equals $p(0)$. This is the minimal Born-rule theorem. A coloured or smooth multiplicative noise generally generates a noise-induced drift when reduced to an It\^o description. The response part of the projected Adler--Millard kernel must cancel that drift while preserving norm, no-signalling at the averaged level and complete positivity where applicable. Work on deterministic coarse-grained localisation supplies useful models but not yet this full microscopic derivation \cite{Kakade2023,Bassi2013}.

The theorem is necessary but not sufficient for a field theory. A many-branch process requires every branch weight to be a bounded martingale on the probability simplex, with absorbing faces and almost-sure completion. A relativistic completion must also specify covariant collapse generators, compatible spacelike composition and an averaged dynamics that does not permit superluminal signalling. These conditions are additional tests, not consequences of the two-branch equation.

\begin{cautionbox}{A necessary separation}
$\cD_{\rm FDT}>0$ says that a projected channel is out of equilibrium. Equation~\eqref{eq:martingale} says that the resulting branch weights have the correct probability law. These are different statements. A viable GTD collapse mechanism must satisfy both, together with norm preservation and completion.
\end{cautionbox}

\subsection{Objective collapse as effective many-body stochastic dynamics}
\label{sec:collapse_bridge}

Phenomenological collapse models provide a useful benchmark for the GTD coarse-graining problem, but they must not be identified with its microscopic solution. In one common convention, the noise-averaged mass-proportional continuous-spontaneous-localisation (CSL) equation is
\begin{equation}
\frac{\dd\rho}{\dd t}
=-\frac{\ii}{\hbar}[H,\rho]
-\frac{\lambda}{2m_0^2}\int\dd^3x\,
\bigl[\widehat M_{r_C}(\bm x),
[\widehat M_{r_C}(\bm x),\rho]\bigr],
\label{eq:csl_benchmark}
\end{equation}
where $\widehat M_{r_C}$ is the mass density smeared over a correlation length $r_C$, $m_0$ is a reference mass and convention-dependent smearing factors may be absorbed into $\lambda$ \cite{GhirardiPearleRimini1990,Bassi2013}. For two approximately classical mass profiles $\mu_1$ and $\mu_2$, Eq.~\eqref{eq:csl_benchmark} suppresses their coherence at a rate of the schematic form
\begin{equation}
\Gamma_{12}
=\frac{\lambda}{2m_0^2}\int\dd^3x\,
\left[\mu_{1,r_C}(\bm x)-\mu_{2,r_C}(\bm x)\right]^2.
\label{eq:csl_amplification}
\end{equation}
The important condensed-matter fact is that amplification is controlled by the spatial density contrast and the material geometry at the resolution $r_C$, not by an undifferentiated count of constituents. A collective state can therefore be macroscopically distinct in current, phase or magnetic moment while having a comparatively small direct mass-density collapse rate.

Equation~\eqref{eq:csl_benchmark} is a target language, not an output already obtained from GTD. A microscopic derivation would have to supply the chain
\begin{equation}
\boxed{
\delta C_{\AS}^{\rm sub}
\ \longrightarrow\ 
\bigl(N_{\rm GTD},K^R_{\rm GTD}\bigr)
\ \longrightarrow\ 
\bigl(\widehat{\cO},\mathcal K_{\rm col},\mathcal K_{\rm diss}\bigr)
\ \longrightarrow\ 
\text{Born-compatible stochastic dynamics}.
}
\label{eq:gtd_to_collapse}
\end{equation}
Here $\widehat{\cO}$ is the physical collapse operator and $\mathcal K_{\rm col},\mathcal K_{\rm diss}$ are its noise and response kernels. The derivation must also determine their normalisation, spatial and temporal correlations, amplification law, subsystem dependence and conversion from Connes time to physical time. In particular, $\widehat{\cO}$ must not simply be declared equal to the CSL mass density.

There is, however, a structural GTD reason to investigate a fermionic or bifermionic $\widehat{\cO}$. For the STM-atom action analysed in Ref.~\cite{SinghFermionCollapse2026}, the purely bosonic subsector has a self-adjoint Hamiltonian, whereas unequal odd-grade Grassmann elements in the fermionic sector generate an intrinsic anti-self-adjoint contribution. This identifies the fermionic sector as the candidate collapse channel within that construction. It does \emph{not} by itself derive Eq.~\eqref{eq:csl_benchmark}, prove mass proportionality, or establish the many-branch Born martingale. Those remain separate coarse-graining tests.

Once a local geometry and the physical-time map exist, condensed-matter response theory provides the operational second half of the bridge. For the derived material operator $\widehat{\cO}_{\bm q}$ define
\begin{equation}
S_{\cO}(\bm q,\omega)
=\int_{-\infty}^{\infty}\dd t\,
\ee^{\ii\omega t}
\left\langle
\widehat{\cO}_{\bm q}(t)\widehat{\cO}_{-\bm q}(0)
\right\rangle .
\label{eq:structure_factor}
\end{equation}
Up to operator normalisation, form factors and detailed-balance conventions, the collapse-induced power deposited in a material is then schematically
\begin{equation}
P_{\rm col}
\sim\int\dd^3q\int_0^\infty\dd\omega\,
\hbar\omega\,
\widetilde N_{\rm GTD}(\bm q,\omega)
S_{\cO}(\bm q,\omega).
\label{eq:collapse_power}
\end{equation}
This formula turns a material into a spectrometer for the GTD fluctuation kernel. Depending on $\widehat{\cO}$ and the support of $\widetilde N_{\rm GTD}$, the response can appear as phonon heating, centre-of-mass force noise, diffusion, quasiparticle production, condensate depletion or loss of interferometric visibility. The dependence of bulk heating on the collapse-noise frequency spectrum and the longitudinal phonon dispersion is explicit in existing CSL analyses \cite{AdlerVinante2018}.

Two restrictions are essential. First, Eqs.~\eqref{eq:structure_factor}--\eqref{eq:collapse_power} use physical $\bm q$, physical $\omega$ and a local material Hamiltonian. Applying them directly to the pre-spacetime $0+1$ dimensional matrix ensemble would be circular; they become legitimate only after Eq.~\eqref{eq:geom_acceptance} and the clock map of Sec.~\ref{sec:scales} have been established. Second, white, nondissipative CSL injects energy without compensating friction and resembles an effectively infinite-temperature bath. A microscopic GTD theory should derive the relation, or controlled violation of the relation, between $\mathcal K_{\rm col}$ and $\mathcal K_{\rm diss}$ rather than importing white noise by hand.

Finally, standard GRW or CSL dynamics is not normally a thermodynamic phase transition. It requires neither a critical temperature nor a large-system nonanalyticity; varying $\lambda$ generally produces a localisation crossover. Collapse becomes a genuine ordering transition in the present framework only if a specified large-system limit satisfies Eq.~\eqref{eq:zero_mode}, or if the coexistence and multi-channel jump conditions in Eq.~\eqref{eq:first_order_coexistence} hold. The branch overlap $q_{\loc}$ remains useful in either case: it can be nonzero branch by branch even when the noise-averaged one-point function preserves the original symmetry. Thus the precise proposal is conditional:
\begin{equation}
\boxed{
\begin{aligned}
&\text{CSL may be to nonequilibrium GTD what a Langevin equation is}\\
&\text{to microscopic many-body dynamics.}
\end{aligned}
}
\label{eq:csl_langevin}
\end{equation}
Establishing this relation requires deriving every arrow in Eq.~\eqref{eq:gtd_to_collapse}; resemblance of equations is not sufficient.

\section{Critical dynamics and universality}
\label{sec:critical}

\subsection{Why ``Model C'' is provisional}

After a local spatial description exists, nonconserved order parameters coupled to conserved energy density resemble Model C in the Hohenberg--Halperin classification \cite{HohenbergHalperin1977}. The GTD composites also couple to the conserved trace Hamiltonian and Adler--Millard charge. Nevertheless, assigning Model C before geometry forms would be premature. The microscopic problem is a many-matrix system with one evolution parameter, nonlocal index structure and constraints. Its large-$N$ or spectral critical behaviour may not fall into a standard finite-dimensional universality class.

The appropriate sequence is:
\begin{enumerate}
\item compute the $0+1$ dimensional matrix susceptibility and dynamic kernel;
\item identify a phase with relational or spectral locality;
\item coarse-grain that phase into spatial fields;
\item only then classify the infrared dynamical universality class.
\end{enumerate}

\subsection{Continuous transition and Kibble--Zurek scaling}

If the post-soldering transition is continuous, define
\begin{equation}
\epsilon=\frac{g-g_c}{g_c},
\qquad
\xi=\xi_0|\epsilon|^{-\nu},
\qquad
\tau_{\rm rel}=\tau_0|\epsilon|^{-\nu z}.
\label{eq:critical_scaling}
\end{equation}
For an approximately linear quench, $\epsilon\simeq \tau/\tau_Q$, adiabaticity fails at
\begin{equation}
|\widehat\epsilon|
=\left(\frac{\tau_0}{\tau_Q}\right)^{1/(1+\nu z)},
\qquad
\widehat\xi
=\xi_0\left(\frac{\tau_Q}{\tau_0}\right)^{\nu/(1+\nu z)}.
\label{eq:KZ}
\end{equation}
This is the Kibble--Zurek freeze-out length \cite{Kibble1976,Zurek1985,Dziarmaga2010}. It could control domains of the leaf selector, time orientation, finite-internal vacuum or discrete exchange symmetry. However, standard Kibble--Zurek reasoning presupposes a causal medium, spatial correlation length and quench clock. It therefore applies directly only to a secondary transition after a proto-geometric phase exists. For the primary birth of spacetime, Eq.~\eqref{eq:KZ} is merely an analogy until the matrix model derives a relational distance, causal propagation bound and clock; before then one should speak only of a matrix-index or spectral correlation range.

\subsection{First-order transition}

If the enlarged sector has two locally stable minima separated by a barrier, the relevant quantities are the bounce action, nucleation rate, latent heat, wall velocity and inverse duration $\beta_{\rm PT}$. A cosmological first-order event near the electroweak temperature can source gravitational waves. A useful redshift relation for a characteristic source frequency $f_*$ is
\begin{equation}
f_0\simeq1.65\times10^{-5}\,\mathrm{Hz}
\left(\frac{f_*}{H_*}\right)
\left(\frac{T_*}{100\,\mathrm{GeV}}\right)
\left(\frac{g_*}{100}\right)^{1/6}.
\label{eq:GWfreq}
\end{equation}
For $f_*/H_*\sim10^2$ and $T_*\sim160$ GeV, the characteristic frequency is of order millihertz. This is a scale estimate, not a prediction of a signal. The amplitude requires a real first-order transition with sufficient released energy and anisotropic stress; the Standard Model crossover does not provide it \cite{Caprini2020}.

Equation~\eqref{eq:GWfreq} also presupposes a semiclassical expanding spacetime during nucleation, propagation and redshift. It may therefore describe a first-order rearrangement \emph{within} an already geometric phase, but not the primary creation of spacetime unless a matching calculation derives an intermediate background on which bubbles and tensor waves are well defined.

\subsection{Crossover}

A crossover has no divergent correlation length, no universal defect density and no bubble nucleation. It is diagnosed by susceptibility peaks and rapid but analytic changes in observables. If the full GTD event is only a crossover, claims of critical exponents, topological defects or first-order gravitational waves must be withdrawn. Localisation could still be a dynamical crossover, but its completion and Born law would require a separate mechanism.

\section{Gravity as the hydrodynamics of the ordered phase}
\label{sec:hydro}

\subsection{Derivative expansion}

Once Eqs.~\eqref{eq:geom_acceptance} hold and all gapped microscopic modes have been integrated out, symmetry permits a long-wavelength action of the form
\begin{align}
\Gamma_{\rm ord}
=\int\dd^4x\sqrt{-g}\bigg[
&-\Lambda_{\eff}+\frac{M_{\Pl}^2}{2}R
+\mathcal L_{\rm matter}\nonumber\\
&+c_1R^2+c_2R_{\mu\nu}R^{\mu\nu}
+c_3R_{\mu\nu\rho\sigma}R^{\mu\nu\rho\sigma}
+\cdots\bigg]
+\Gamma_{\rm noise/diss}.
\label{eq:IRaction}
\end{align}
The Einstein equation is then the leading constitutive relation, in the same broad sense in which hydrodynamics organises the leading derivative response of a material. Jacobson's equation-of-state derivation is a close conceptual precedent \cite{Jacobson1995}. The spectral action supplies a separate route to the allowed low-energy operator classes \cite{ChamseddineConnes1997,SinghUniverse2026}.

The analogy has limits. Analogue-gravity systems show how effective causal metrics and horizon kinematics can arise, but they do not by themselves derive the full Einstein dynamics \cite{Volovik2003,Barcelo2011}. A graviton is not literally an ordinary sound mode, and diffeomorphism redundancy, local Lorentz symmetry, the equivalence principle and a universal spin-two coupling must be derived or protected. Calling gravity ``hydrodynamics'' does not make these requirements disappear. Nor does it prohibit quantised collective fluctuations: phonons are quantised even though elasticity is emergent.

\subsection{Newton's constant as stiffness}

Let $h^{\rm TT}_{ij}$ be the transverse-traceless response around a selected background. With a specified normalisation, define the geometric stiffness from the static inverse two-point function,
\begin{equation}
\Gamma^{(2)}_{\rm TT}(0,\bm k)
=\mathcal Z_g M_{\Pl}^2\bm k^2+O(k^4),
\qquad
M_{\Pl}^2=\frac{1}{\mathcal Z_g}
\left.\frac{\partial\Gamma^{(2)}_{\rm TT}(0,\bm k)}{\partial\bm k^2}\right|_{\bm k=0}.
\label{eq:stiffness}
\end{equation}
Here $\mathcal Z_g$ is fixed by the convention for $h^{\rm TT}$. Matching $M_{\Pl}^2=(8\pi G)^{-1}$ defines Newton's constant as a response coefficient. In the present programme, the Planck scale already appears at the one-aikyon level. The electroweak-neighbourhood event is proposed to select and match the phase; it is not asserted to generate the enormous hierarchy $M_{\Pl}/v$.

The definition must be implemented in a background-field construction that fixes gauge, projects onto the physical transverse--traceless sector and proves the emergent Ward identities. The static and long-wavelength limits need not commute; the order $\omega\to0$ followed by $\bm k\to0$ must be stated. The crucial microscopic calculation is a Kubo-type relation expressing Eq.~\eqref{eq:stiffness} in terms of connected GTD stress, co-frame or Dirac-spectrum correlators. If the response contains additional unsuppressed scalar or vector poles, the equivalence-principle and fifth-force constraints become immediate failure tests.

There is also an explicit two-scale problem. Numerically,
\begin{equation}
\frac{M_{\Pl}^2}{v^2}\sim10^{32},
\label{eq:two_scale}
\end{equation}
so identifying the ordering neighbourhood with the electroweak scale does not explain the gravitational stiffness. A viable derivation must show whether this ratio comes from a protected microscopic coefficient, a large collective factor, critical scaling or another calculable mechanism, without inserting it as a fitted Landau parameter.

Finally, the hydrodynamic analogy brings a dissipation question. The observed vacuum propagates gravitational waves with extremely small attenuation. For a retarded TT kernel one may define
\begin{equation}
\eta_g=-\lim_{\omega\to0}\frac{1}{\omega}
\operatorname{Im}K^R_{\rm TT}(\omega,\bm k=0).
\label{eq:grav_diss}
\end{equation}
The ordered vacuum must yield vanishing or observationally negligible $\eta_g$ in the relevant regime, even if dissipative kernels were important during formation. General relativity would then describe the nearly nondissipative infrared constitutive limit.

\subsection{Four routes to Einstein dynamics: induced, constrained, thermodynamic and holographic}
\label{sec:four_routes}

Sakharov, constrained $BF$ gravity, Jacobson and Padmanabhan are often grouped under ``emergent gravity'', but they answer different questions and begin at different logical levels. Treating them as interchangeable would obscure precisely what the present programme must derive. Sakharov explains how an Einstein--Hilbert coefficient can be induced; constrained $BF$/Plebanski theory explains how Einstein gravity is selected as a classical branch of a gauge theory; Jacobson explains why the Einstein equation can be an equilibrium equation of state; and Padmanabhan develops the entropy, heat-content and holographic structure of the already emergent spacetime. The four accounts can be complementary only if their normalisations and domains of validity match.

\paragraph{Sakharov: an induced effective action.}
Sakharov starts with quantum fields on a smooth background geometry and integrates out their fluctuations. In modern heat-kernel language the result has the schematic form \cite{Sakharov1968,Visser2002}
\begin{equation}
\Gamma_{\rm 1\mbox{-}loop}[g]
=\int\dd^4x\sqrt{-g}\left[
c_0\Lambda_{\rm UV}^4
{}+c_1\Lambda_{\rm UV}^2R
{}+c_2\ln\!\left(\frac{\Lambda_{\rm UV}}{\mu}\right)R^2
{}+\cdots\right].
\label{eq:sakharov_action}
\end{equation}
Thus vacuum polarisation supplies metrical elasticity, with $M_{\Pl,\rm ind}^2\sim c_1\Lambda_{\rm UV}^2$. This is a genuine effective-action calculation, but it presupposes the differentiable metric, matter-field content, regulator and quantum state. It does not explain the birth of the metric or select a Lorentzian four-dimensional branch. It also induces a cosmological term and higher-curvature terms, so recovering the observed hierarchy is not automatic.

The GTD analogue is therefore not necessarily ``pure induced gravity''. If the broken $BF$ sector already gives a tree-level Einstein coefficient while the surviving quantum fields add Sakharov-type corrections, the renormalised stiffness has the form
\begin{equation}
M_{\Pl,\rm ren}^2
=M_{\Pl,BF}^2+\Delta M_{\Pl,\rm ind}^2+\Delta M_{\Pl,\rm micro}^2.
\label{eq:planck_matching}
\end{equation}
Equation~\eqref{eq:stiffness} measures this one renormalised coefficient; it is not an additional contribution to be added again. A complete matching calculation must prevent double counting between the microscopic action, the constrained-$BF$ branch and the quantum determinant.

\paragraph{Wesley--Singh--Isidro: the Einstein--Hilbert branch of broken $BF$ theory.}
The $SO(3,3)$ construction of Wesley, Singh and Isidro supplies a more direct action-level bridge than the thermodynamic analogy by itself \cite{WesleySinghIsidro2026}. Its parent six-dimensional, split-signature gauge theory is broken to chiral $SU(2)_L\times SU(2)_R$ sectors; components of the coset connection furnish effective co-frames, and the gravitational sector is written as a constrained chiral $BF$ theory. In schematic Plebanski notation,
\begin{equation}
S_{\rm Pl}[B,A,\Psi]
=\frac{1}{8\pi G_{BF}}\int\left[
B^i\wedge F_i(A)
-\frac12\left(\Psi_{ij}+\frac{\Lambda_{BF}}{3}\delta_{ij}\right)
B^i\wedge B^j\right].
\label{eq:plebanski_action}
\end{equation}
Variation of the traceless multiplier $\Psi_{ij}$ gives the simplicity equation already used in Eq.~\eqref{eq:simplicity}. On its non-degenerate branch,
\begin{equation}
B^i=P^i{}_{ab}\,e^a\wedge e^b,
\qquad \det(e^a{}_{\mu})\neq0,
\label{eq:bf_simple_solution}
\end{equation}
with appropriate Lorentzian reality conditions. Substitution into Eq.~\eqref{eq:plebanski_action} gives the chiral Palatini form and hence the Einstein--Hilbert action with a cosmological term, up to convention-dependent topological, boundary and chiral completion terms:
\begin{equation}
S_{\rm Pl}\big|_{\rm simple}
\longrightarrow
\frac{1}{16\pi G_{BF}}
\int\epsilon_{abcd}\,e^a\wedge e^b\wedge
\left(R^{cd}-\frac{\Lambda_{BF}}{6}e^c\wedge e^d\right)+\cdots.
\label{eq:bf_to_eh}
\end{equation}

The same mechanism is visible in MacDowell--Mansouri form. Once the broken connection has
\begin{equation}
F^{ab}=R^{ab}+\ell^{-2}e^a\wedge e^b,
\label{eq:mm_curvature}
\end{equation}
the projected curvature square contains
\begin{equation}
\epsilon_{abcd}F^{ab}\wedge F^{cd}
=\epsilon_{abcd}\left[
R^{ab}\wedge R^{cd}
+\frac{2}{\ell^2}e^a\wedge e^b\wedge R^{cd}
+\frac{1}{\ell^4}e^a\wedge e^b\wedge e^c\wedge e^d
\right].
\label{eq:mm_expansion}
\end{equation}
The middle term is Einstein--Hilbert, the last is cosmological, and the first is topological in four dimensions under the usual assumptions. This is the precise sense in which the $BF$ paper derives Einstein gravity from the underlying gauge action.

The word ``derives'' nevertheless has a bounded meaning. The result is a classical low-energy or broken-branch derivation conditional on the $SO(3,3)$ breaking, dimensional reduction to the two four-dimensional leaves, the identification of the coset connection with co-frame data, the simplicity and non-degeneracy equations, Lorentzian reality, interface consistency and the gravitational branch choice. It does not by itself derive the GTD statistical transition, objective localisation, the Born rule, the horizon area law or a microscopic state count. Conversely, the present GTD paper should not merely reimpose the $BF$ simplicity constraint after postulating an unrelated co-frame. The decisive bridge is to show that the soft geometric eigenmode of the many-aikyon susceptibility produces the compensator/coset and $B$ fields entering Eqs.~\eqref{eq:plebanski_action}--\eqref{eq:mm_curvature}.

\paragraph{Jacobson: Einstein's equation as a local equation of state.}
Jacobson begins downstream of both the GTD ordering problem and the $BF$ branch selection. Assuming a Lorentzian spacetime with local Rindler horizons, the Unruh temperature, an entropy density $\eta$ per unit horizon area and local equilibrium, he imposes \cite{Jacobson1995}
\begin{equation}
\delta Q=T\,\dd S,
\qquad
T=\frac{\hbar\kappa}{2\pi},
\qquad
\dd S=\eta\,\dd A.
\label{eq:jacobson_clausius}
\end{equation}
Raychaudhuri focusing, validity for every local null direction and stress-energy conservation then yield
\begin{equation}
G_{ab}+\Lambda g_{ab}=8\pi G_{\rm hor}T_{ab},
\qquad
G_{\rm hor}=\frac{1}{4\hbar\eta}.
\label{eq:jacobson_G}
\end{equation}
This explains the constitutive or equation-of-state character of the Einstein equation but assumes the causal horizon, local Lorentz invariance, area entropy and equilibrium state whose microscopic origin GTD seeks. For more general entropy densities, an internal entropy-production term can be required \cite{ElingGuedensJacobson2006}. The GTD noise and retarded kernels are candidates for a microscopic source of such dissipation only after a horizon coarse graining has been constructed; a nonzero fluctuation--dissipation defect cannot simply be renamed $\dd_iS$.

\paragraph{Padmanabhan: entropy extremisation and holographic evolution.}
Padmanabhan emphasises that the field equations of diffeomorphism-invariant gravity can be expressed as local heat/entropy identities, and develops variational principles in which null directions rather than the metric are varied \cite{Padmanabhan2009,Padmanabhan2010}. He also interprets cosmological expansion as evolution toward holographic equipartition \cite{Padmanabhan2012},
\begin{equation}
\frac{\dd V}{\dd t}
=L_{\Pl}^2\bigl(N_{\rm sur}-N_{\rm bulk}\bigr),
\qquad
N_{\rm sur}=N_{\rm bulk}\quad\text{in de~Sitter equilibrium}.
\label{eq:padma_equipartition}
\end{equation}
This is the broadest thermodynamic architecture of the four approaches: gravitational dynamics is encoded in horizon heat content, Noether entropy, surface/bulk degrees of freedom and their disequilibrium. It is not, by itself, a microscopic enumeration of spacetime atoms. For GTD, $N_{\rm sur}$, $N_{\rm bulk}$ and the null-surface entropy must be computed from the ordered ensemble rather than assigned by relabelling aikyons.

\begin{table}[htbp]
\centering
\caption{Four distinct routes to Einstein dynamics and the layer still missing from each.}
\label{tab:four_routes}
\small
\begin{tabularx}{\textwidth}{@{}p{0.17\textwidth}p{0.24\textwidth}YY@{}}
\toprule
Route & Starting structure & Principal output & Not supplied by that route\\
\midrule
Sakharov & Matter QFT on a smooth background metric & Induced $R$, cosmological and higher-curvature terms & Birth and branch selection of spacetime; universal finite coefficients without a regulator/microscopic completion\\
Constrained $BF$ & Broken gauge connection, two-forms, simplicity and reality conditions & Einstein--Hilbert/Palatini action on the non-degenerate gravitational branch & Statistical origin of the branch, horizon state count and Born/localisation dynamics\\
Jacobson & Local causal horizons, Unruh temperature, area entropy and local equilibrium & Einstein equation as a local equation of state & Microscopic origin of the metric, area law and equilibrium\\
Padmanabhan & Null-surface entropy/heat density and holographic degree counts & Field equations as thermodynamic identities and cosmic expansion toward equipartition & A specified microscopic ensemble and derivation of the degree counts\\
Present GTD programme & Pre-geometric matrix dynamics and many-aikyon ensemble & Candidate common ordering/localisation mechanism and calculable response kernels & The regulated calculation that actually yields the $BF$ branch, $S_{\rm EH}$ and horizon entropy\\
\bottomrule
\end{tabularx}
\end{table}

The synthesis can now be stated as a stringent matching problem rather than a family resemblance. Let $G_{BF}$ be extracted from Eq.~\eqref{eq:bf_to_eh}, $G_{\rm Kubo}$ from the TT stiffness in Eq.~\eqref{eq:stiffness}, and $G_{\rm hor}$ from the microscopic area-entropy density in Eq.~\eqref{eq:jacobson_G}. A single-substrate completion requires, after common renormalisation and convention matching,
\begin{equation}
\boxed{
G_{BF}=G_{\rm Kubo}=G_{\rm hor},
\qquad
\eta_{\rm GTD}=\frac{1}{4\hbar G_{BF}}.}
\label{eq:three_G_matching}
\end{equation}
The $BF$ route would then tell us which action governs the ordered branch; Sakharov would describe quantum renormalisation by the modes that remain; Jacobson would recover the local equilibrium equation of state; and Padmanabhan would encode its entropy and cosmological heat balance. Failure of Eq.~\eqref{eq:three_G_matching} would show that these are not yet limits of one microscopic theory, even if each separately reproduces Einstein equations.

\subsection{Verlinde's elastic gravity and the cosmological infrared action}
\label{sec:verlinde}

Verlinde's emergent-gravity construction is the closest phenomenological precedent for the present stiff--elastic reading of gravity \cite{Verlinde2017}. In that proposal the positive dark-energy contribution changes an otherwise ``stiff'' spacetime geometry into a medium with a slow, elastic response. Localised matter removes or displaces part of the de~Sitter entropy; the microscopic state is argued to be glassy and to retain a memory of that displacement. The residual strain then produces an additional, MOND-like gravitational force at an acceleration of order $cH_0$. This is more than a verbal similarity: both approaches connect de~Sitter structure, matter localisation, memory and a nonlinear infrared gravitational response.

The comparison nevertheless requires a phase-language correction. Verlinde's ``stiff geometry'', ``dark elastic phase'' and glassy memory are macroscopic response descriptions. His paper does not exhibit an order parameter, a thermodynamic large-system limit or a singular inverse susceptibility separating general relativity from the dark elastic regime. It also does not derive the response from a controlled microscopic Hamiltonian or ensemble. The GR-to-MOND change should therefore be called a constitutive crossover unless a nonanalyticity is independently demonstrated. This is exactly the distinction made in Sec.~\ref{sec:hydro}: the possible genuine transition is
\begin{equation}
\text{pre-geometric GTD ensemble}
\longrightarrow
\text{ordered classical geometry},
\label{eq:primary_geom_transition}
\end{equation}
whereas GR and MOND can be two response limits within the same ordered phase.

The cosmological sector of the $E_8\times\omega E_8$ programme supplies an explicit variational carrier for this second crossover \cite{SinghCosmology2026}. Let the ordered vacuum define a foliation scalar $T(x)$, with unit timelike normal and acceleration
\begin{equation}
u_\mu=-\frac{\nabla_\mu T}{\sqrt{-g^{\alpha\beta}\nabla_\alpha T\nabla_\beta T}},
\qquad
a_\mu=u^\nu\nabla_\nu u_\mu,
\qquad
\mathcal I[g,T]=a_\mu a^\mu.
\label{eq:foliation_acceleration}
\end{equation}
In units $c=1$, the proposed infrared functional is
\begin{equation}
S_{\rm IR}[g,T]
=\frac{a_0^2}{16\pi G}\int\dd^4x\sqrt{-g}\,
F\!\left(\frac{\mathcal I[g,T]}{a_0^2}\right),
\label{eq:SIR}
\end{equation}
with a matching counterterm understood so that the high-acceleration quadratic term reproduces rather than double-counts the Einstein weak-field action. Its required asymptotics are
\begin{equation}
F(y)\sim y\quad(y\gg1),
\qquad
F(y)\sim\frac{2}{3}y^{3/2}\quad(y\ll1).
\label{eq:F_asymptotics}
\end{equation}
The prefactor in Eq.~\eqref{eq:SIR} cannot by itself be identified with the AQUAL normalisation, because the metric variation also contains the Einstein--Hilbert term and $\Delta S_{\rm match}$. The required matching condition is that their complete quasistatic weak-field reduction, including the matter coupling and up to the overall action-sign convention, be
\begin{equation}
S_{\rm qs}[\Phi]
=-\frac{a_0^2}{8\pi G}\int\dd t\,\dd^3x\,
F\!\left(\frac{|\bm\nabla\Phi|^2}{a_0^2}\right)
-\int\dd t\,\dd^3x\,\rho\Phi.
\label{eq:AQUAL_matching_action}
\end{equation}
Variation of Eq.~\eqref{eq:AQUAL_matching_action} then gives the AQUAL equation \cite{BekensteinMilgrom1984}
\begin{equation}
\bm\nabla\!\cdot\!\left[
\mu\!\left(\frac{|\bm\nabla\Phi|}{a_0}\right)
\bm\nabla\Phi\right]=4\pi G\rho,
\qquad
\mu(x)=F'(x^2).
\label{eq:AQUAL_from_SIR}
\end{equation}
Thus $\mu\rightarrow1$ at high acceleration, while $\mu\sim x$ in the deep infrared and spherical solutions obey $g\simeq\sqrt{a_0g_N}$. The quantity
\begin{equation}
K_{\rm tan}(x)=\frac{\dd(\mu g)}{\dd g}=\mu(x)+x\mu'(x)
\label{eq:tangent_modulus}
\end{equation}
is a literal tangent modulus for the nonlinear gravitational constitutive law. Equations~\eqref{eq:SIR}--\eqref{eq:tangent_modulus} therefore make the required stiff--elastic matching operational rather than deriving it: the programme must still show explicitly that $\Gamma_{\rm EH}+S_{\rm IR}+\Delta S_{\rm match}$ reduces to Eq.~\eqref{eq:AQUAL_matching_action}, with neither a factor-of-two mismatch nor an additional high-acceleration contribution to $\mu$.

The de~Sitter relation is also sharper in this effective description. Restoring $c$,
\begin{equation}
a_0=\frac{c^2}{\xi\ell_{\rm dS}},
\qquad
\Lambda_{\eff}=\frac{3}{\ell_{\rm dS}^2},
\label{eq:a0Lambda}
\end{equation}
so the background curvature scale and the infrared response scale have a common vacuum origin. Set $F(0)=0$, or absorb its constant part into $\Lambda_{\eff}$. With that convention $S_{\rm IR}$ is inert on an exactly homogeneous FRW background because $a_\mu=0$ there. The cosmological constant drives the background de~Sitter expansion; $S_{\rm IR}$ governs inhomogeneous scalar response and bound systems. De~Sitter is therefore the substrate of the infrared scale, not itself the MOND force law.

\begin{table}[htbp]
\centering
\caption{The structural overlap and the non-equivalence of Verlinde's construction and the GTD programme.}
\label{tab:verlinde}
\small
\begin{tabularx}{\textwidth}{@{}p{0.19\textwidth}YY@{}}
\toprule
Question & Verlinde & Present programme\\
\midrule
Microscopic substrate & Unspecified de~Sitter microstates with area/volume-law entanglement, slow relaxation and memory & Deterministic many-aikyon matrix dynamics and a regulated ensemble, still to be solved\\
Matter localisation & Localised matter displaces entropy carried by delocalised dark-energy excitations & Objective fermionic or bifermionic localisation is to follow from projected Adler--Millard fluctuations\\
Ordinary gravity & Otherwise stiff spacetime response & Ordered-phase TT stiffness $M_{\Pl}^2$ defined by Eq.~\eqref{eq:stiffness}\\
Infrared response & Elastic memory gives an apparent dark force at $a_0\sim cH_0$ & $S_{\rm IR}[g,T]$ yields GR and MOND asymptotics with $a_0$ tied to $\ell_{\rm dS}$\\
Missing derivation & No controlled micro-to-elastic calculation or complete relativistic cosmology & No derivation yet of $S_{\rm IR}$, $F$, $\xi$ or the foliation dynamics from GTD correlators\\
\bottomrule
\end{tabularx}
\end{table}

The strongest possible synthesis is consequently a hypothesis, not a result:
\begin{equation}
\boxed{
\text{many-aikyon correlators}
\stackrel{?}{\longrightarrow}
\Gamma_{\rm IR}[g,T]
=\Gamma_{\rm EH}+S_{\rm IR}+\Delta S_{\rm match}.}
\label{eq:micro_to_SIR}
\end{equation}
If $\mathcal A_i$ denotes the microscopic composite projected onto the emergent acceleration channel, the required calculation has the Kubo form
\begin{equation}
\frac{\delta^2\Gamma_{\rm IR}}
{\delta a_i\,\delta a_j}
\stackrel{?}{=}
\left[\chi^R_{\mathcal A_i\mathcal A_j}(0,\bm k)\right]^{-1},
\label{eq:SIR_Kubo}
\end{equation}
with Ward identities, order of limits, stability and matching fixed. It must explain why the kernel crosses from $F(y)\sim y$ to $F(y)\sim(2/3)y^{3/2}$ and must derive the normalisation in Eq.~\eqref{eq:a0Lambda}. Until that is done, GTD supplies a candidate substrate and the cosmological paper supplies an action-level closure, but the arrow between them remains open. The advantage over a purely entropic construction is therefore not that the microphysics has already been solved; it is that the missing micro-to-constitutive calculation can be stated in terms of a definite ensemble and response kernel.

\subsection{A two-component interpretation}

It is useful to picture a localised geometric component coexisting with an unlocalised quantum component. Define, only after a dynamically meaningful localisation projector $P_{\loc}$ exists,
\begin{equation}
f_{\loc}=\frac{\Tr(P_{\loc}\rho^{(1)})}{\Tr\rho^{(1)}},
\qquad
f_{\rm q}=1-f_{\loc}.
\label{eq:fractions}
\end{equation}
This avoids identifying unrelated cosmological counts with condensate fractions. The split is not literally the Landau two-fluid model of helium: the ``ordered component'' supplies geometry and classical sources, while the normal component supplies quantum degrees of freedom. Conservation laws and energy exchange between the components must be derived from the many-aikyon theory.

\section{Three scales and the clock-matching problem}
\label{sec:scales}

The following quantities are conceptually distinct:
\begin{equation}
\beta_{\TD}^{-1},
\qquad
T_{\rm cosmological},
\qquad
\mu_{\rm RG}.
\label{eq:three_scales}
\end{equation}
$\beta_{\TD}$ is the Lagrange multiplier conjugate to the trace Hamiltonian. A cosmological temperature presupposes an emergent local clock and thermal matter. The renormalisation scale is a bookkeeping scale for running couplings. Numerical proximity does not identify them.

Suppose the ordered co-frame defines proper time $t$ along a comoving trajectory and, near a stationary background,
\begin{equation}
\dd t=Z_t\,\dd\tau.
\label{eq:clockmap}
\end{equation}
Frequencies obey $\omega_{\tau}=Z_t\omega_t$. Matching the KMS exponent gives
\begin{equation}
\beta_{\rm phys}=Z_t\beta_{\TD},
\qquad
T_{\rm phys}=\frac{T_{\TD}}{Z_t}
\label{eq:temperaturemap}
\end{equation}
in units $k_B=1$. The response coefficient $Z_t$ must be computed from the same ordered phase; before Eq.~\eqref{eq:clockmap} exists, ``the GTD temperature is 160 GeV'' has no invariant meaning.

A natural matching scale in the broken phase might be the inverse correlation length, an amplitude-mode gap, or a threshold eigenvalue of $D_F$,
\begin{equation}
\mu_*\sim \xi^{-1},\quad m_{\rm amp},\quad\text{or}\quad |\lambda(D_F)|,
\label{eq:mustar}
\end{equation}
but the choice must follow from a matching calculation. The observed electroweak crossover is a datum against which the derived physical temperature can be compared, not an input to $\beta_{\TD}$.

\section{Connection to the finite Dirac and flavour problem}
\label{sec:flavour}

The critical-emergence programme may supply the missing vacuum-selection law of the finite Dirac operator. This is attractive because the same bifermionic sector that seeds scalar/internal fields can, in principle, choose chiral frames and phases. It is also tightly constrained by the recent CKM audit \cite{SinghCKM2026}.

Integrating out a heavy complementary block gives the low-energy Schur complement
\begin{equation}
Y_f^{\eff}=A_f-P_fC_f^{-1}Q_f.
\label{eq:schur}
\end{equation}
Self-adjointness of the full grading-odd $D_F$ relates the reverse-chiral block to $Y_f^{\dagger}$; it does not require $Y_f$ to be Hermitian or the left and right complement planes to coincide. Therefore the general condensate should initially retain independent row and column structures.

An economical Hermitian ansatz would take
\begin{equation}
Y_f^{\eff}=A_f-\Phi K_f\Phi^{\dagger},
\qquad
\Phi\in\mathbb C^{3\times2},\qquad K_f=K_f^{\dagger},
\label{eq:rank2}
\end{equation}
with one common two-plane for the up and down sectors. If $\Delta_f=A_f-Y_f^{\eff}$, an exact common plane requires a nonzero vector $n$ such that
\begin{equation}
\Delta_un=\Delta_dn=0.
\label{eq:common_null}
\end{equation}
Under the canonical identification $A_f=\mathcal Q_f$ with the diagonal Jordan modulus, Hermitian left--right locking, common-scale quark spectra and the measured CKM target, a global optimisation over the common $SU(3)$ locking frame gives
\begin{equation}
\min_{S\in SU(3)}
\sigma_{\min}
\begin{pmatrix}\Delta_u(S)\\\Delta_d(S)\end{pmatrix}
=1.03\times10^{-4}
\label{eq:rank2defect}
\end{equation}
on the top-normalised scale. The defect is about fourteen times $m_u/m_t$ in the same input set. Independent global searches converge to the same nonzero minimum. This is a conditional numerical no-go for the conjunction
\begin{equation}
A_f=\mathcal Q_f,
\qquad \text{Hermitian locking},
\qquad \text{one common rank-two plane}.
\label{eq:no_go_assumptions}
\end{equation}
It is not a no-go for a full matrix condensate, a biunitary bridge, three directions, sector-dependent projections of one master field, or additional derived Peirce terms in $A_f$. The condensed-matter lesson is precise: do not truncate the family order parameter before the microscopic susceptibility identifies its soft subspace.

Because Eq.~\eqref{eq:rank2defect} is new numerical work rather than a quoted result, the deposit bundle includes the optimisation code, inputs, machine-readable results and the dedicated technical note. The no-go should be reassessed if the quark inputs, scale convention or direct-block hypothesis changes; it must not be promoted to a theorem beyond the assumptions in Eq.~\eqref{eq:no_go_assumptions}.

\section{Possible empirical windows}
\label{sec:phenom}

The framework is useful only if it eventually produces observables. Table~\ref{tab:windows} separates robust conditional implications from speculative possibilities.

\begin{table}[htbp]
\centering
\caption{Phenomenological windows and the calculations required before they become predictions.}
\label{tab:windows}
\small
\begin{tabularx}{\textwidth}{@{}p{0.23\textwidth}YY@{}}
\toprule
Window & Required microscopic output & Interpretation\\
\midrule
Collapse heating, diffusion or loss of visibility & Normalised $N_{AB}(\omega)$, physical collapse operator, amplification law and clock conversion & Direct test of the projected nonequilibrium kernel; existing collapse bounds apply only after this map\\
Material-resolved collapse response & $\widehat{\cO}$, $\widetilde N_{\rm GTD}(\bm q,\omega)$, response kernel and form factors & Phonons, force noise, quasiparticles and condensates probe complementary regions of $S_{\cO}(\bm q,\omega)$\\
Frequency structure of collapse noise & Poles or lines of the Adler--Millard correlator in physical frequency units & Tests proposed GTD noise models; spectral positions cannot be quoted before $Z_t$ is fixed\\
Infrared gravity and MOND response & Acceleration-channel susceptibility, derived $F(y)$, $a_0$--$\Lambda$ normalisation, foliation stability and relativistic perturbations & Distinguishes a GTD-derived constitutive law from an inserted AQUAL closure; galaxy, lensing and cosmological tests then probe the same kernel\\
Cosmological defects & Vacuum manifold, homotopy, order and quench rate & Kibble--Zurek scaling applies only for a continuous transition with emergent locality\\
Stochastic gravitational waves & First-order bounce, $\alpha$, $\beta_{\rm PT}/H_*$, wall velocity and source lifetime & Electroweak-neighbourhood first-order events can lie in the mHz band; no signal follows from a crossover\\
Extra gravitational polarisations or fifth forces & Complete pole decomposition of the geometric response & Strong equivalence-principle and binary constraints; an unsuppressed extra pole is a likely falsifier\\
Running couplings and spectral-action relations & Derived cutoff moments, finite traces and threshold matching & Must be evaluated at one consistent scheme and scale; fitted support factors are not predictions\\
CKM/PMNS data & Vacuum eigenvector and full left/right finite-Dirac projections & Out-of-sample flavour test after the condensate is fixed in one sector\\
\bottomrule
\end{tabularx}
\end{table}

The most distinctive near-term theoretical observable is not yet a number. It is the pole and spectral-projector structure of the matrix susceptibility $\chi_{AB}(\omega)$, supplemented by an eigenvector only for a nondegenerate pole. It decides whether there is one transition, several transitions, a crossover, or no relevant instability. Once the time map is known, the same kernel gives relaxation rates and noise spectra. A wider catalogue of proposed experimental tests is given in Ref.~\cite{SinghPredictions2026}; the present framework states which additional kernels are needed before the transition-related entries become predictions.

\section{Red-team analysis: where the analogy can fail}
\label{sec:redteam}

The condensed-matter analogy is productive precisely because it supplies failure tests.

\begin{enumerate}
\item \textbf{No normalisable ensemble.} If the regulated phase-space integral cannot be made convergent without destroying the Adler--Millard symmetry or the desired continuum limit, the statistical foundation fails.

\item \textbf{No collective instability.} The susceptibilities may remain finite, or their soft mode may have support only in a Higgs or localisation channel. Then a common emergence event is false.

\item \textbf{Localisation without geometry.} A nonzero branch overlap and small commutators may coexist with zero four-volume, failed simplicity, wrong signature or spectral dimension other than four.

\item \textbf{Geometry without universal coupling.} An ordered co-frame may form while matter species couple with different effective metrics. The equivalence principle would then fail unless violations are sufficiently suppressed.

\item \textbf{Wrong low-energy poles.} Extra gapless scalars, vectors, ghosts or acausal modes in the ordered response would conflict with observation or unitarity.

\item \textbf{Born-rule failure.} A non-unitary fluctuation can localise states but yield a biased drift, non-completion, signalling or loss of norm. Localisation alone is not enough.

\item \textbf{Imported collapse phenomenology.} If the GTD collapse operator, spatial kernel and physical-time map are not derived, applying CSL mass-density bounds or a material structure factor merely assumes the missing bridge. Conversely, identifying standard CSL with a phase transition would confuse stochastic localisation with thermodynamic nonanalyticity.

\item \textbf{Imported infrared elasticity.} Similarity to Verlinde's de~Sitter response does not derive $S_{\rm IR}$. If $F(y)$, the $a_0$--$\Lambda$ normalisation, the matching counterterm and stable foliation dynamics cannot be obtained from ordered-phase correlators, MOND remains a separate effective closure rather than an output of GTD.

\item \textbf{Scale conflation.} Fitting $\beta_{\TD}^{-1}$ directly to $159$ GeV without deriving $Z_t$ would remove the quantitative content of the electroweak claim.

\item \textbf{Unexplained stiffness hierarchy.} A common ordering event near the electroweak scale does not explain $M_{\Pl}^2/v^2\sim10^{32}$. If this ratio is inserted into an effective coefficient, gravity has been matched but not derived.

\item \textbf{Conflating action and thermodynamic derivations.} The constrained-$BF$ construction can yield the Einstein--Hilbert action without deriving horizon state counting, while the Jacobson--Padmanabhan relations can recover field equations or heat identities without selecting the $BF$ branch. Treating either result as completion of the other skips a physical layer.

\item \textbf{Newton-coefficient mismatch.} The coefficient $G_{BF}$ in the broken action, the static $G_{\rm Kubo}$ measured by the ordered response and the $G_{\rm hor}$ inferred from area entropy may disagree after renormalisation. Then the proposed microscopic, action and thermodynamic descriptions cannot yet represent one infrared theory.

\item \textbf{Gauge-dependent order parameters.} A claimed Higgs transition based only on $\langle H\rangle$ in a fixed gauge does not establish a gauge-invariant thermodynamic event.

\item \textbf{Tautological potential.} If the coefficients of $V_{\eff}$ are chosen from CKM, Higgs and gravitational data, the construction has parametrised those data rather than explained them.

\item \textbf{Circular cosmology.} Standard quench lengths, bubbles, Hubble rates and wave redshifts assume a causal spacetime. Applying them to the primary creation of spacetime without a proto-geometric matching regime is circular.

\item \textbf{Emergent spin-two obstruction.} The Weinberg--Witten theorem constrains massless composite spin-two particles in Lorentz-covariant theories with a suitable conserved stress tensor \cite{WeinbergWitten1980}. The programme may evade its assumptions because the microscopic theory is pre-spacetime and diffeomorphism invariance is emergent, but that evasion must be shown, not asserted.
\end{enumerate}

\section{Prioritised open programme}
\label{sec:roadmap}

Table~\ref{tab:roadmap} turns the analogy into a sequence of calculations. The ordering is deliberate. Later phenomenology is not meaningful if the earlier measure and kernel do not exist.

\begin{longtable}{@{}p{0.055\textwidth}p{0.25\textwidth}p{0.36\textwidth}p{0.25\textwidth}@{}}
\caption{Open calculations, their pass criteria and what failure would mean.}\label{tab:roadmap}\\
\toprule
No. & Calculation & Decisive deliverable & Meaning of failure\\
\midrule
\endfirsthead
\toprule
No. & Calculation & Decisive deliverable & Meaning of failure\\
\midrule
\endhead
1 & Regulated many-aikyon measure & A positive or controlled contour measure, finite $Z$, Liouville property, conserved constraints and a stated large-regulator limit & No foundation for equilibrium or susceptibility calculations\\
2 & Composite operator basis & Regulator-compatible, Grassmann-even, gauge-covariant/invariant operators for localisation, geometry, Higgs and finite-internal channels & The proposed order parameters are not observables of the microscopic theory\\
3 & Static susceptibility matrix & $\chi_{AB}(0)$ with finite-size scaling and its soft spectral projectors (and eigenvector when nondegenerate) across control parameters & No evidence for a common instability; possibly several unrelated transitions\\
4 & Master-field test & A microscopic representation showing whether the four channels are projections of one composite & Coincidence requires multicritical tuning or a first-order mechanism\\
5 & Real-time kernel & Retarded response, symmetrised noise, KMS test and $\cD_{\rm FDT}$ for the projected Adler--Millard channel & No quantitative route from equilibrium fluctuations to localisation\\
6 & Collapse map, Born and norm audit & Derived $\widehat{\cO}$, spatial/frequency kernel, amplification law and physical-time map, together with zero branch-weight drift, completion, norm preservation and acceptable averaged dynamics & CSL comparison is imported rather than derived, or collapse fails even if states become sharply peaked\\
7 & Geometric reconstruction & A microscopic map to the Wesley--Singh--Isidro compensator/coset and constrained-$BF$ variables; simplicity, nonzero four-volume, Lorentzian reality, $d_s\to4$, locality and a consistent spin structure & Localisation does not select the non-degenerate Einstein branch of the proposed gauge theory\\
8 & Clock and scale map & $Z_t$, physical KMS temperature and matching rule for $\mu_*$ & Electroweak-scale coincidence remains dimensionally and operationally undefined\\
9 & Ordered-phase response and thermodynamic matching & Kubo value of $M_{\Pl}^2$, origin of $M_{\Pl}^2/v^2$, TT dissipation bound, full pole content and universal matter coupling; microscopic area entropy satisfying $G_{BF}=G_{\rm Kubo}=G_{\rm hor}$; acceleration-channel derivation of $S_{\rm IR}$ and $F(y)$ & Gravity is not recovered, the action/response/entropy coefficients disagree, precision constraints fail, or the MOND closure remains phenomenological\\
10 & Order and kinetics & Continuous exponents, first-order bounce data, or a demonstrated crossover; then a proto-geometric causal regime before quench/domain predictions & Kibble--Zurek or gravitational-wave claims are inapplicable or circular\\
11 & Finite-Dirac vacuum selection & Full matrix condensate, chiral projection, right-frame dynamics and predicted CKM/PMNS outputs & Flavour remains a moduli-space fit; the common rank-two shortcut is already disfavoured\\
12 & One-scale phenomenological audit & Collapse, gravity, electroweak and flavour quantities in consistent conventions with propagated uncertainties & Quantitative incompatibility falsifies the proposed common phase\\
\bottomrule
\end{longtable}

\subsection{The first feasible calculation}

The most realistic first calculation is a finite regulated ensemble with a small number of aikyons and a finite auxiliary Grassmann algebra. One should:
\begin{enumerate}
\item preserve global unitary invariance exactly;
\item measure the Adler--Millard charge and the proposed even composites;
\item at every finite size establish stationarity, estimate integrated
autocorrelation or mixing times and take
$\Delta\tau\gg\tau_{\rm corr}$ before interpreting ensemble averages;
\item scan a dimensionless control parameter such as $\beta_{\TD}$ times a regulated energy scale;
\item compute the full connected covariance matrix and Binder cumulants;
\item perform finite-size scaling and test whether the soft spectral projector
and channel weights are stable as matrix size and aikyon number increase;
\item compute real-time or Connes-time response by perturbing with small sources;
\item compare the measured symmetrised correlator with the appropriate classical fluctuation--response relation or, after quantum emergence, the quantum KMS prediction.
\end{enumerate}
This finite model would not prove the continuum theory. Long trajectories
would test temporal self-averaging and measure finite-system correlators;
variation of aikyon number and matrix size would test collective scaling. Only
the second operation can establish a genuine critical singularity. Together
they answer a more basic question: does the microscopic action possess even a
candidate collective instability with the required channel content?

\subsection{A first execution in a surrogate ensemble: selected diagnostics are calculable}
\label{sec:toyexec}

Selected calculations proposed above have now been executed once, in a simple
surrogate that respects the requirements needed for those diagnostics while
remaining fully computable: exact global unitary invariance, a conserved
bosonic Adler--Millard charge, an even multi-channel composite vector,
noncommutative-geometric ordering behaviour with branch discontinuities, and a
matter sector dragged by that ordering. The ensemble is \emph{not} the GTD
measure; the purpose of this subsection is to demonstrate that the central
objects of Secs.~\ref{sec:micro}--\ref{sec:noneq} are calculable and that the
acceptance criteria behave as designed on data, not to claim evidence for GTD
itself. All results are conditional on the stated surrogate. The complete
pipeline (code, raw data, figures) is included in the ancillary bundle; the
published data are one single-provenance run of the final code with
deterministic seeds, produced after earlier exploratory runs --- used only to
develop the sampler guards described below --- were discarded in full.

\paragraph{The surrogate ensemble.}
Three traceless Hermitian $N\times N$ matrices $Q_a$ carry the geometric
channel through the Myers (Yang--Mills--Chern--Simons) action, whose
first-order transition between a disordered ``matrix'' phase and a condensed
fuzzy-sphere phase, $Q_a\simeq\phi L_a$ with $L_a$ the
spin-$\frac{N-1}{2}$ $su(2)$ generators, is established in the matrix-model
literature \cite{AzumaBalNagaoNishimura2004,OConnorYdri2013}. One Hermitian
matrix $M$ carries the matter channel. The regulated action is the
single-trace invariant
\begin{equation}
S=N\Tr\!\Big[-\tfrac14\textstyle\sum_{ab}[Q_a,Q_b]^2
+\tfrac{2\ii\alpha}{3}\epsilon_{abc}Q_aQ_bQ_c
+\tfrac{\rho}{2N}\textstyle\sum_aQ_a^2
+\tfrac{m^2}{2}M^2+\tfrac{g_4}{4}M^4
+\tfrac{\tilde\kappa}{2N}M^2\textstyle\sum_aQ_a^2\Big],
\label{eq:toy_action}
\end{equation}
with $\alpha=\tilde\alpha/\sqrt N$ and, throughout,
$(\rho,m^2,g_4,\tilde\kappa)=(0.5,\,1,\,0.5,\,1.5)$. The $\rho/N$ term is a
declared regulator: it lifts and confines the commuting flat directions of the
matrix phase while shifting the sphere-existence condition only at $O(1)$
($\tilde\alpha^2>2\rho$). The $\tilde\kappa/N$ scaling makes the matter
back-reaction on the geometry and the geometric shift of the matter mass both
$O(1)$ at large $N$. On an irreducible fuzzy sphere, or under the mean-field
replacement $N^{-1}\sum_aQ_a^2\simeq o_1\mathbf1$, the latter reduces to the
induced-mass mechanism of Eq.~\eqref{eq:induced_masses}, with
$m^2_{\rm eff}=m^2+\tilde\kappa\,o_1$. On a general matrix configuration the
shift is instead the matrix-valued anticommutator operator generated by
$\frac{\tilde\kappa}{2N}M^2\sum_aQ_a^2$.
Sampling is by hybrid Monte Carlo whose molecular-dynamics segments are
literally trace-dynamics trajectories of the trace Hamiltonian
$H=\tfrac12\Tr P^2+S$ at $\beta=1$; matrix sizes are $N=8,12,16,24$ with
cold- and fuzzy-sphere-start branches. The composite vector is
$\bm\Phi=(o_1,\dots,o_5)$ with
$o_1=\Tr\sum_aQ_a^2/N^2$ (geometric amplitude),
$o_2=-\ii\,\epsilon_{abc}\Tr Q_aQ_bQ_c/N^2$ (geometric orientation),
$o_3=-\sum_{a<b}\Tr[Q_a,Q_b]^2/N^2$ (noncommutativity),
$o_4=\Tr M^2/N$ (matter) and $o_5=\Tr M^2\!\sum_aQ_a^2/N^2$ (cross channel);
all are even, invariant words in the fundamental matrices, as
Sec.~\ref{sec:orders} requires. The sphere-start collapse threshold,
$\tilde\alpha\simeq2.4$--$2.6$ across the studied $N$, sits consistently
above the pure-Myers lower spinodal $(8/3)^{3/4}\!\approx\!2.09$
\cite{AzumaBalNagaoNishimura2004}, the displacement being the expected
regulator and matter back-reaction shift.

\paragraph{Discrete Adler--Millard conservation.}
The bosonic Adler--Millard charge $\widetilde C=\sum_a[Q_a,P_a]+[M,P_M]$ is
conserved along the leapfrog trajectories to machine precision (relative
in-trajectory drift $\sim8\times10^{-16}$, independent of step size). This is
stronger than the naive $O(\dd\tau^2)$ symplectic-integrator expectation, and
the reason is worth recording: the leapfrog kick preserves $\widetilde C$
because global unitary invariance of $S$ gives the Noether identity
$\sum_r[q_r,\partial S/\partial q_r]=0$ \emph{identically}, and the drift
preserves it because $[p_r,p_r]=0$. (The trace projection that keeps the
$Q_a$ traceless adds a multiple of the identity to the force and drops out of
every commutator.) The discretised flow therefore inherits the exact
conservation law of continuum trace dynamics. The property is not automatic
for arbitrary schemes: it requires kicks built from exact gradients of an
invariant action and force-free drifts; preconditioned or
momentum-dependent-kernel integrators can preserve the invariant measure
without conserving $\widetilde C$ exactly. Within that (standard) integrator
class, charge drift is a clean diagnostic of a broken implementation rather
than of discretisation error.

\begin{figure}[htbp]
\centering
\includegraphics[width=\textwidth]{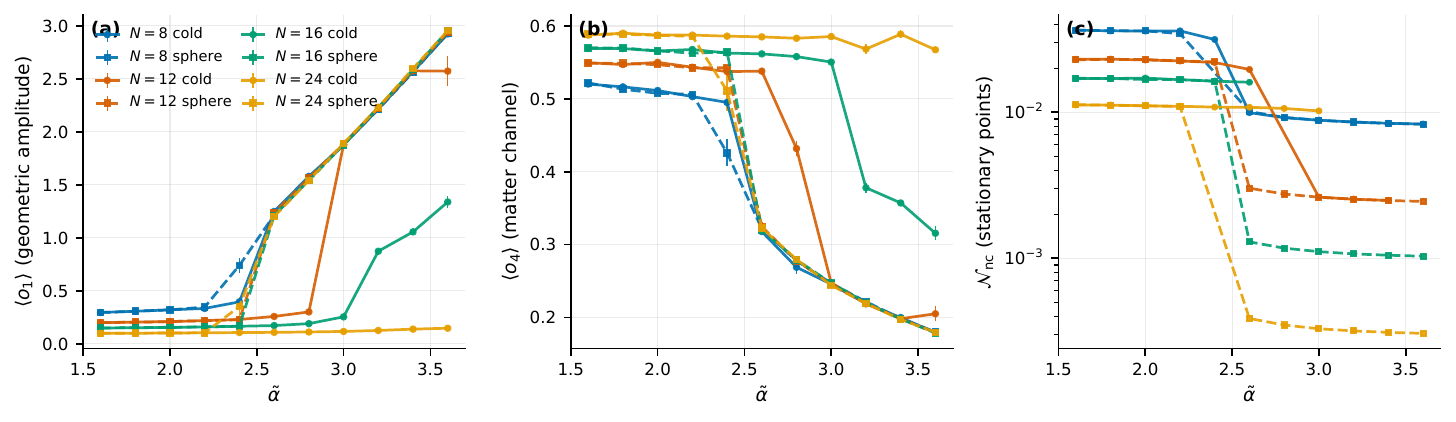}
\caption{Evidence consistent with first-order multi-channel ordering in the surrogate ensemble
($\tilde\kappa=1.5$). (a)~Geometric amplitude $\langle o_1\rangle$ for cold
(solid) and fuzzy-sphere (dashed) starts: the jump is discontinuous and the
metastability window widens with $N$; at $N=8$ the branches merge (ergodic
tunnelling at $\tilde\alpha\simeq2.4$), while at $N=24$ the cold branch
remains disordered across the whole scan. Cold-start nucleation at
intermediate $N$ is strongly seed-dependent and, at $N=16$, proceeds through
long-lived partial configurations; only branch-level statements (existence
and widening of the window, inter-branch discontinuities) are used in the
text, never the precise nucleation locations, and low-acceptance points
(Fig.~\ref{fig:toy_chi}a, open symbols) carry unquantified non-ergodicity
systematics. (b)~The matter channel $\langle o_4\rangle$ is dragged
discontinuously at each branch's transition --- the induced-mass mechanism of
Eq.~\eqref{eq:induced_masses}. In the $\tilde\kappa=0$ control the same
quantity is flat at $0.631(2)$ across the entire scan. (c)~Normalised
noncommutativity on stationary points: the ordered phase is an order of
magnitude more nearly commuting per unit amplitude.}
\label{fig:toy_phase}
\end{figure}

\begin{figure}[htbp]
\centering
\includegraphics[width=\textwidth]{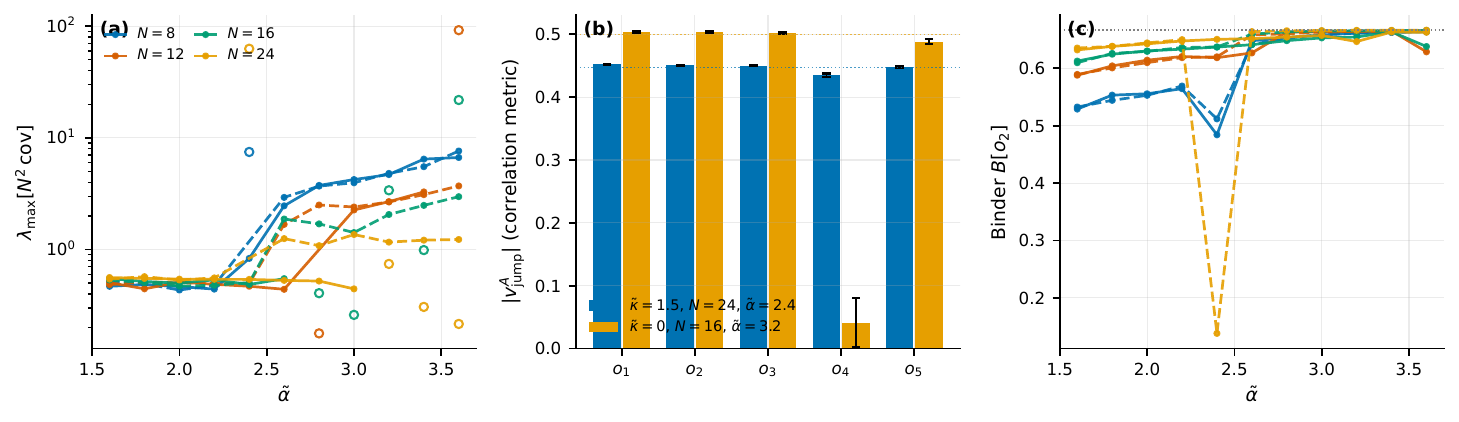}
\caption{Stationary branch covariances and the transient jump-direction null test.
(a)~$\lambda_{\max}[\chi_{AB}(0)]$: $O(1)$ with no $N$-growth on stationary
branch points (filled; classification criteria in the text), spiking at
nucleation/tunnelling/low-acceptance points (open) --- behaviour consistent
with a first-order jump, with no observed continuous softening on either
sampled stationary branch. (b)~Leading principal direction of the transient
correlation matrix (the
operator metric $G_{AB}$ fixed by unit diagonal covariance) at a
transition transient of the coupled ensemble versus the decoupled control, with
jackknife errors; dotted lines mark the rank-one reference values $1/\sqrt5$
and $1/2$ (see text). The matter-channel support is $0.435(3)$ when the
coupling exists and $0.04(4)$ when it does not. (c)~Binder cumulant of
$o_2$: dips at transition transients; away from them it approaches
$2/3$ in the ordered phase (and $0.48$--$0.66$ in the disordered phase).}
\label{fig:toy_chi}
\end{figure}

\paragraph{Phase structure: evidence consistent with a first-order multi-channel jump.}
Scanning $\tilde\alpha\in[1.6,3.6]$ from both starts
(Fig.~\ref{fig:toy_phase}) yields pronounced hysteresis and branch
discontinuities of the kind expected for the known Myers first-order
transition, now deformed by the regulator and matter coupling. Throughout, a scan point is
classified as a \emph{stationary} single-branch measurement by four
$\lambda$-independent conditions --- minority-basin occupancy below $5\%$ in
its stored $o_1$ time series, absence of split-half drift, a small jackknife
error on $o_1$, and Metropolis acceptance $\ge0.5$. Every masked point is
masked by the basin, drift or acceptance condition, never by the error
condition alone. Within the sampled stationary branches no softening trend is
seen: the leading
eigenvalue of
$\chi_{AB}=N^2\langle\delta o_A\delta o_B\rangle_c$ stays $O(1)$ (range
$0.43$--$7.6$) with no increase in $N$ --- if anything it decreases at fixed
$\tilde\alpha$. This scan therefore supplies no evidence for the continuous
zero-mode criterion of Eq.~\eqref{eq:zero_mode}. At fixed
$\tilde\alpha=2.6$, $N=16$,
the two locally stable branches differ by $\Delta o_1=+1.036(8)$,
$\Delta o_2=+0.602(6)$, $\Delta o_3=+0.362(4)$ and $\Delta o_4=-0.242(3)$ (a
$43\%$ matter-channel discontinuity), so $\Delta\bm\Phi$ has nonzero
invariant projection in every channel simultaneously. The finite-size
signatures are consistent with first-order kinetics: $\lambda_{\max}$ spikes by up to two orders of
magnitude at the nucleating or tunnelling points
(Fig.~\ref{fig:toy_chi}a, open symbols), and the Binder cumulant of $o_2$
dips sharply at the same points (Fig.~\ref{fig:toy_chi}c). Because phase
free energies or equal-weight histograms were not computed, these observations
do not establish the exact coexistence condition of
Eq.~\eqref{eq:first_order_coexistence}. The normalised noncommutativity drops by a
factor $\simeq12$ across the jump ($0.0163\to0.0013$ on the $N=16$ sphere
branch), the classicalisation direction anticipated in
Sec.~\ref{sec:orders}.

\paragraph{Transient jump-direction support: a null-test calibration that passes.}
Because raw composites carry unequal normalisations, the leading transient
principal direction is extracted from the correlation matrix, i.e.\ with an
operator metric $G_{AB}$ fixed by unit diagonal
covariance. The honest framing of this test is a \emph{calibration}, not a
discovery: the matter drag is put into the action by $\tilde\kappa$, and at
$\tilde\kappa=0$ the matter sector factorises exactly, so zero matter
support is guaranteed a priori. What the calculation verifies is that the
estimator, at finite statistics and with the metric convention fixed as the
framework requires, returns the built-in answer on both sides. It does: at
the $N=24$ sphere-branch nucleation transient the leading direction is
$|v^A_{\rm jump}|=(0.452,0.451,0.450,0.435,0.448)$, with jackknife errors of
$0.001$--$0.003$ --- all five channels participating, close to the uniform
rank-one value $1/\sqrt5\simeq0.447$ that a single collective slow mode
shared by five channels must produce --- whereas the $\tilde\kappa=0$ control
at its own transition gives $(0.503,0.504,0.503,0.041,0.488)$, i.e.\ the
four-channel rank-one value $1/2$ with matter support $0.04(4)$, consistent
with the finite-statistics floor (neighbouring control transients give
$\le0.10$, with the caveat that jackknifed moduli of noise components are
positive-biased). Such a transient correlation matrix is dominated by drift
along the branch-jump vector and is not an equilibrium susceptibility or a
Gaussian critical eigenspace. Its informative content is limited to which
channels participate in the observed jump; the test
confirms that the estimator identifies that set correctly and that the
$\tilde\kappa=0$ null comes out null. A future GTD calculation would face
the nontrivial equilibrium version of this test, where the channel content is
not chosen by hand and the critical spectral projector is obtained from
stationary finite-size scaling.

\begin{figure}[htbp]
\centering
\includegraphics[width=\textwidth]{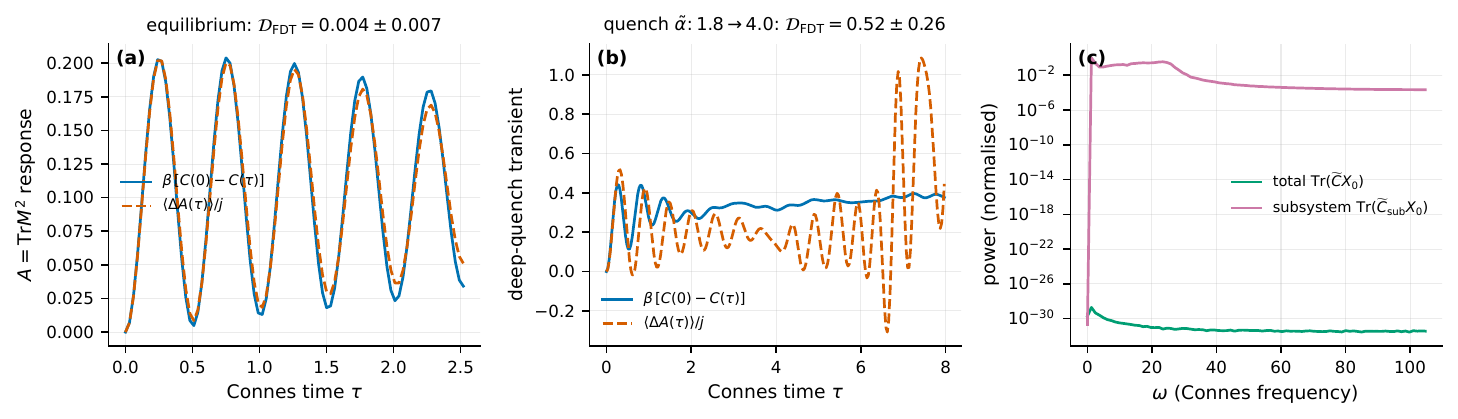}
\caption{The real-time sector at $N=12$.
(a)~Classical equilibrium Kubo test for $A=\Tr M^2$ on the sphere branch at
$\tilde\alpha=2.6$: the step response $\langle\Delta A(\tau)\rangle/j$ lies
on the classical fluctuation--response prediction $\beta[C(0)-C(\tau)]$ through five oscillation periods
of the coherent matter mode; $\cD_{\rm FDT}=0.004\pm0.007$ (jackknife over
configurations). (b)~Deep quench $\tilde\alpha\!:1.8\to4.0$ from the
disordered phase, window $T=16$: $\cD_{\rm FDT}=0.52\pm0.26$ while
$\langle A\rangle$ drifts from $6.60$ to $5.33$ (ordering in progress). The
matched-window shallow quench to $\tilde\alpha=3.2$ gives
$\cD_{\rm FDT}=0.28\pm0.19$ with $\langle A\rangle$ nearly flat
($6.60\to6.49$): its defect is dominated by the onset of nucleation in a
minority of configurations, and over the early window $\tau\le5$ it is only
$0.06$, versus $0.23$ for the deep quench. (c)~Fluctuation power spectra:
the total conserved charge functional carries its (conserved) value entirely
at zero frequency --- its fluctuation spectrum, after mean subtraction, sits
at the round-off floor, $\sim\!29$ orders below the signal --- while the
subsystem charge
$\widetilde C_{\rm sub}=\sum_r[\Pi q_r\Pi,\Pi p_r\Pi]$ fluctuates with
$O(1)$ relative amplitude and a broadband spectrum.}
\label{fig:toy_fdt}
\end{figure}

\paragraph{Fluctuation--dissipation: equilibrium pass, quench violation, and
the projection requirement.}
The classical real-time programme corresponding to the low-frequency limit of
Sec.~\ref{sec:noneq} was executed at $N=12$
(Fig.~\ref{fig:toy_fdt}). The equilibrium classical step-response Kubo relation on the
sphere branch is verified with a time-domain, band-limited defect
$\cD_{\rm FDT}=0.004\pm0.007$ --- the discretised analogue of
Eq.~\eqref{eq:DFDT}, zero within its jackknife error. Two quenches from the
disordered phase at $\tilde\alpha=1.8$, in matched windows ($T=16$,
identical step size), then show the defect doing its intended job: a deep
quench to $\tilde\alpha=4.0$, where ordering proceeds visibly within the
window, gives $\cD_{\rm FDT}=0.52\pm0.26$; a shallow quench to
$\tilde\alpha=3.2$, where the disordered configuration is long-lived and
$\langle A\rangle$ barely drifts, gives a small early-window defect ($0.06$
for $\tau\le5$, versus $0.23$ for the deep quench in the same window),
growing to $0.28\pm0.19$ only as nucleation begins in a minority of
configurations --- with a configuration-to-configuration spread that
dominates the error. The defect therefore responds to genuine
disequilibrium (ordering in progress), remains small while a long-lived
branch stays locally stationary, and is zero in equilibrium --- the qualitative
behaviour required of it in the localisation programme, established here
with modest statistics and correspondingly large errors on the quench
values. Finally, the projection statement of Sec.~\ref{sec:noneq} is
exhibited sharply: linear functionals of the conserved total charge are
constant along every trajectory (relative variation $8\times10^{-16}$),
while the same functionals of the subsystem's own charge
$\widetilde C_{\rm sub}$, built from block-truncated variables, fluctuate
with $O(1)$ relative amplitude and a broadband spectrum
(Fig.~\ref{fig:toy_fdt}c). Broadband ``collapse noise'' can live only in a
subsystem-projected channel; the conserved total cannot supply it even in
principle, exactly as argued below Eq.~\eqref{eq:DFDT}.

\begin{cautionbox}{What the toy execution does and does not establish}
It establishes the calculability of selected diagnostics: a regulated invariant
measure, charge conserved to machine precision, even composite vector, stationary branch
covariances, a transient jump-direction estimator with a fixed operator
metric, multi-channel branch discontinuities with induced-mass drag, a
classical equilibrium Kubo/FDT pass, a quench defect, and the
conserved-versus-subsystem noise dichotomy have been computed once in a
regulated unitary-invariant matrix ensemble on commodity hardware. It does
not determine a phase free-energy crossing, an exact coexistence point or an
equilibrium critical eigenspace. It establishes nothing
about GTD itself: the surrogate of this subsection is bosonic (for the
Grassmann-regulated extension see Sec.~\ref{sec:toyferm}), has no
split-bioctonionic structure and no claim on the GTD measure; its
underlying Myers sector is known to have a first-order transition and the
deformed scan is consistent with that behaviour; the matter drag and
hence the multi-channel support result are equally by construction, so the
transient support test functions here as a null-test calibration of the estimator,
not as a discovery; and the fuzzy-sphere phase realises only branch freezing
and falling $\cN_{\rm nc}$, not simplicity, Lorentzian reality or
$d_s\to4$. The exercise also yields practical warnings for future GTD
simulations, recorded in the ancillary documentation: unregulated flat
directions make the disordered phase a runaway, and near-spinodal stiffness
lets unstable integrator trajectories amplify round-off into configurations
(including non-Hermitian ones, whose Yang--Mills action is unbounded below)
that a bare Metropolis step will accept --- finiteness, physical-region and
Hermiticity guards are mandatory, and their necessity was discovered the
hard way in this calculation. Item~1 of Table~\ref{tab:roadmap} --- the
regulated \emph{GTD} measure --- remains the bottleneck; what this subsection
removes is any doubt that items~3 and~5 are computable once item~1 exists.
\end{cautionbox}

\subsection{Second execution: a Grassmann-regulated fermionic extension}
\label{sec:toyferm}

Two fermionic ingredients of the framework --- a finite Grassmann sector in
the regulated measure and fermionic spectral data as geometric diagnostics
--- have now also been executed once. Two degenerate Grassmann-valued
$N\times N$ matrix-fermion flavours $\psi_f$, $f=1,2$, are coupled covariantly
to the dynamical geometry. Each $\psi_f$ and $\bar\psi_f$ has $2N^2$
spinor--matrix components, so the finite Grassmann algebra contains
$4(2N^2)$ independent generators in total. The Dirac operator and action are
\begin{equation}
\cD_\psi[Q]=\sum_a\sigma_a\otimes\mathrm{ad}_{Q_a}+m_\psi ,
\qquad
S_F=\sum_{f=1}^{2}\bar\psi_f\,\cD_\psi[Q]\,\psi_f ,
\label{eq:toy_dirac}
\end{equation}
acting on the $2N^2$-dimensional (spinor)$\,\otimes\,$(matrix) space, with
$m_\psi=0.2$ throughout. Because $\mathrm{ad}_{Q_a}$ is Hermitian and
transforms covariantly under global $U(N)$, $\cD_\psi$ is Hermitian and
$\det\cD_\psi$ is exactly invariant. The Grassmann integral is performed
\emph{exactly}, giving the positive weight
$[\det\cD_\psi]^2=\det(\cD_\psi^\dagger\cD_\psi)$ and the Grassmann-regulated effective action
$S_{\rm eff}=S_B-\Tr\log(\cD_\psi^\dagger\cD_\psi)$. This is precisely the
``finite auxiliary Grassmann algebra'' requested in the enumeration of
Sec.~\ref{sec:roadmap}, realised at toy scale: in the dynamical runs below ($N=8$) the Grassmann
sector is part of the measure, not a spectator, while at $N=16,24$ it is a
quenched probe of the bosonic ensembles. Two exact zero modes of
$\mathrm{ad}_Q$ (the trace direction) sit at $\lambda=m_\psi$ for every
configuration and are inert.

The construction was validated against exact results before use: for
fuzzy-sphere configurations $Q_a=\phi L_a$ the operator \eqref{eq:toy_dirac}
reduces (up to normalisation and the mass shift) to the
Grosse--Pre\v{s}najder Dirac operator of the fuzzy two-sphere
\cite{GrossePresnajder1995,CarowWatamura1997}, and the computed spectrum reproduces its exact structure ---
integer-spaced levels $\lambda=\phi k$ with linearly growing degeneracies ---
to machine precision, which is the Weyl-law signature of a two-dimensional
geometry. The determinant force was validated against numerical gradients at
$10^{-5}$, and the fermionic force satisfies the Noether identity
$\sum_a[Q_a,F_a^\psi]=0$ at machine precision, as it must, since the
integrated-out Grassmann sector preserves global unitary invariance exactly.

\begin{figure}[htbp]
\centering
\includegraphics[width=\textwidth]{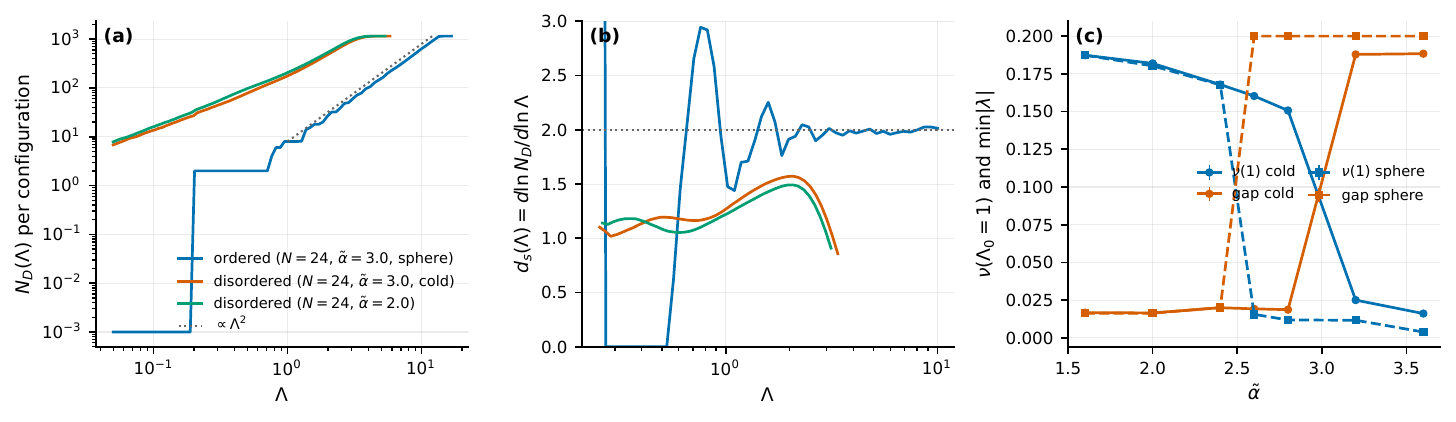}
\caption{Fermionic spectroscopy of the emergent geometry (quenched, exact
Grassmann integration; $m_\psi=0.2$). (a)~Ensemble-averaged counting function
$N_D(\Lambda)$ of $\cD_\psi$ at $N=24$: the ordered phase follows the
two-dimensional Weyl law $N_D\propto\Lambda^2$ over its upper decade (dotted
guide); the step at $\Lambda=m_\psi$ is the pair of exact trace zero-modes.
(b)~Windowed spectral-dimension flow $d_s(\Lambda)$: the ordered ensemble
settles onto a plateau at $d_s=1.96$ (top-decade fit; exact value $2$),
approached through oscillations that reflect the discreteness of the fuzzy
spectrum; the disordered ensembles drift at $1.0$--$1.5$ with no plateau.
(c)~The fermion-derived spectral channel across the transition at $N=16$ (jackknife
errors, mostly smaller than symbols): the mode-number fraction
$\nu(\Lambda_0{=}1)$ (blue) collapses and the spectral gap (orange) opens at
each branch's transition point (sphere branch dashed, cold branch solid);
the cold branch saturates at $0.1879(13)<m_\psi$, resolving its sub-maximal
ordered state.}
\label{fig:toy_ferm1}
\end{figure}

\paragraph{A fermionic geometric diagnostic: the spectral-dimension flow.}
The spectral-dimension entry of the geometric acceptance set,
Eq.~\eqref{eq:weyl}, is now executed on a dynamical ensemble
(Fig.~\ref{fig:toy_ferm1}a,b). At $N=24$ the ensemble-averaged counting
function $N_D(\Lambda)$ of $\cD_\psi$ in the ordered (fuzzy-sphere) phase
follows the two-dimensional Weyl law over its upper decade, with a windowed
log-derivative that settles onto a clean plateau: a top-decade fit gives
$d_s=1.96\pm0.02$ (the error is the fit-window systematic; a per-configuration
statistical error was not stored), against the exact value $2$ for the
emergent two-sphere. The oscillatory approach to the plateau is physical ---
it is the discreteness of the fuzzy spectrum, i.e.\ the short-distance
granularity of the emergent geometry --- and the plateau is reached in the
semiclassical regime. As with the support criterion, the honest framing is a
null test: the ordered phase consists by construction of fluctuations around
a configuration whose exact spectrum this section validates, so
$d_s\simeq2$ there was guaranteed for any unbroken estimator; the nontrivial
content is that the dynamical fluctuations do not destroy the plateau, and
that the disordered ensembles (at $\tilde\alpha=3.0$ and $2.0$) yield
\emph{no} false plateau --- their top-decade slopes drift at $1.28$--$1.36$.
The diagnostic thus identifies the correct dimension when a geometry is
present and refuses to produce one when it is not, which is the behaviour
the framework needs from Eq.~\eqref{eq:weyl} before it can be trusted on GTD
data.

\begin{figure}[htbp]
\centering
\includegraphics[width=0.8\textwidth]{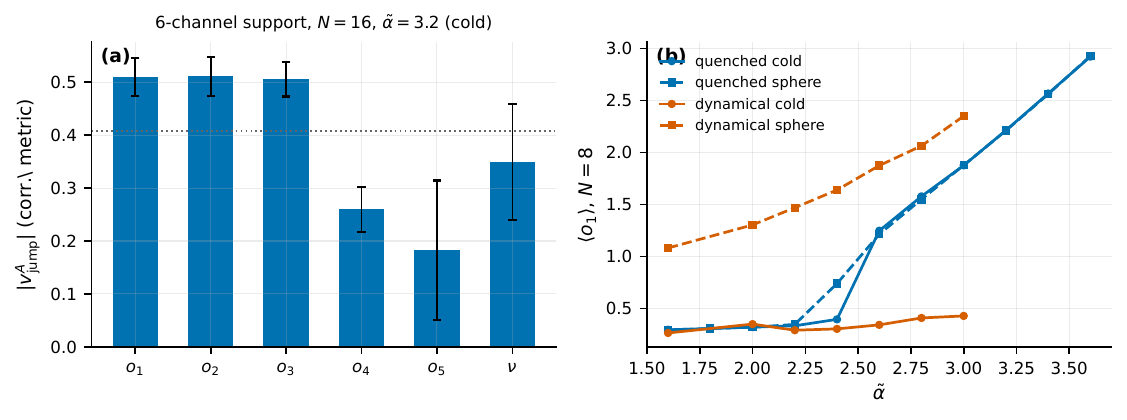}
\caption{(a)~Six-channel leading transient jump direction (correlation metric, jackknife
errors) at the $N=16$ cold-branch nucleation point $\tilde\alpha=3.2$: the
fermion-derived spectral channel $\nu$ participates ($0.35\pm0.11$) alongside the
geometric and matter channels; dotted line: uniform six-channel value
$1/\sqrt6$. (b)~Back-reaction of the dynamical Grassmann sector at $N=8$:
with the exact determinant in the measure (orange) the fuzzy-sphere branch
survives to $\tilde\alpha=1.6$, far below the quenched collapse (blue), and
branch tunnelling is frozen in both directions (determinant-suppressed
nucleation); in a dedicated integrator check the bosonic Adler--Millard
charge is conserved at $10^{-15}$.}
\label{fig:toy_ferm2}
\end{figure}

\paragraph{A fermion-derived spectral channel in the transient covariance matrix.}
The statistically robust fermion-derived scalar is the mode-number
fraction $\nu(\Lambda_0)=\#\{|\lambda|<\Lambda_0\}/2N^2$ at fixed
$\Lambda_0=1$ (the condensate-like $\Tr\cD_\psi^{-1}$ and
$\Tr\cD_\psi^{-2}$ are dominated by rare near-zero modes in the disordered
phase and are unsuitable as covariance channels; this is itself a useful
lesson for the choice of composite basis in roadmap item~2). The mode number
is a gauge-invariant spectral functional of $\cD_\psi$ after the Grassmann
variables are integrated out; it is not itself a bifermionic operator
$\bar\psi K\psi$ and does not execute the localisation composite of
Eq.~\eqref{eq:bilinear}. Across the
geometric transition $\nu$ collapses by an order of magnitude and the
spectral gap opens at each branch's transition point
(Fig.~\ref{fig:toy_ferm1}c): on the $N=16$ sphere branch,
$\nu:0.1677(9)\to0.0156(2)$ and
$\min|\lambda|:0.0201(13)\to0.2000(0)=m_\psi$ between
$\tilde\alpha=2.4$ and $2.6$. The delayed cold-branch nucleation at
$\tilde\alpha=3.2$ shows the analogous but \emph{not identical} jump,
$\nu\to0.0251(3)$ and gap $\to0.1879(13)$ --- nine standard deviations
below $m_\psi$ --- because the cold branch nucleates into a sub-maximal
(partially ordered, likely reducible-representation) configuration rather
than the maximal fuzzy sphere, and the fermionic diagnostics resolve the
difference. The fermionic spectrum is thus dragged by the geometric ordering
along with the matter channel --- gap opening and mode depletion are the
spectral face of the same branch jump --- and, as a bonus, these
fermion-derived observables discriminate between distinct ordered states that
the geometric amplitude alone only hints at. Added as a
sixth channel, $\nu$ appears in the leading transient principal direction of the correlation
matrix with support $0.35\pm0.11$ (Fig.~\ref{fig:toy_ferm2}a), alongside
the geometric ($\simeq0.51$) and matter ($0.26\pm0.04$) channels. Three
disclosures accompany this number. The point shown is selected as the one
with the largest leading correlation eigenvalue (the $N=16$ cold-branch
nucleation point at $\tilde\alpha=3.2$), not by a pre-registered criterion.
The channel is degenerate in the fully gapped phase: there $\nu$ has exactly
zero sample variance (the spectrum is rigid), its correlation row is
undefined, and the estimator returns zero support --- the spectral channel
is informative only near the transition, which itself is a lesson for
composite-basis design. And the error, a jackknife over eight time-ordered
blocks of $\sim100$ autocorrelated samples, is order-of-magnitude only; the
support is three standard deviations from zero but within one of the
six-channel uniform value $1/\sqrt6\simeq0.41$. As in
Sec.~\ref{sec:toyexec}, the participation is by construction ($\cD_\psi$
depends on $Q$) and the statement is a calibration of the transient
jump-direction estimator on a six-channel problem. It is not a measurement
of an equilibrium critical spectral projector or of a bifermionic
localisation condensate.

\paragraph{Dynamical back-reaction: the Grassmann sector in the measure.}
Finally, the Grassmann sector was promoted from probe to participant: at
$N=8$ the scan was repeated with $\det(\cD_\psi^\dagger\cD_\psi)$
included exactly in both the Metropolis test and the molecular-dynamics
force. Three results (Fig.~\ref{fig:toy_ferm2}b). First, in a dedicated
integrator check the bosonic Adler--Millard charge is conserved to machine
precision ($|\Delta\widetilde C|/|\widetilde C|\sim10^{-15}$) along
trace-dynamics trajectories of the enlarged measure, confirming the
integrator-class theorem of Sec.~\ref{sec:toyexec} in the presence of the
fermionic force, whose Noether identity $\sum_a[Q_a,F^\psi_a]=0$ holds
identically. Second, the back-reaction is large: \emph{both} metastability
windows widen dramatically. The fuzzy-sphere branch, which collapses below
$\tilde\alpha\simeq2.4$--$2.6$ in the quenched ensemble, survives fully
gapped down to $\tilde\alpha=1.6$ (where $\langle o_1\rangle=1.083(3)$),
while the disordered branch survives past $\tilde\alpha=3.0$, where the
quenched $N=8$ ensemble has long since ordered. Widening in both directions
is the signature of nucleation suppression rather than of a simple tilt of
the free-energy balance: tunnelling paths must pass through configurations
with near-vanishing Dirac eigenvalues, on which the determinant weight
collapses --- the analogue of determinant-induced tunnelling suppression
familiar from dynamical-fermion lattice simulations. The equal-weight
transition point therefore cannot be located from hysteresis endpoints, and
we do not claim it. Third, at the level of the action itself the Grassmann
sector does favour the gapped branch: at fixed couplings inside the
hysteresis window, the measured branch difference of the fermionic effective
action is $\langle S_{F,\rm eff}\rangle_{\rm sphere}-\langle S_{F,\rm
eff}\rangle_{\rm cold}=-258.6(6)$ at $\tilde\alpha=2.0$ and $-237.1(5)$
at $2.4$ --- an action-level preference of order $N^2$ for the gapped phase.
This is not a free-energy determination (entropic contributions are not
measured), so the two observations together say: the determinant pushes the
action balance toward the gapped geometry while simultaneously freezing the
sampler's ability to realise that preference. For a future GTD simulation
the warning is concrete: a fermionic sector in the measure can qualitatively
reshape both the phase diagram and the ergodics of the sampler.

\begin{cautionbox}{Scope of the fermionic extension}
What is established: the regulated measure can carry an exactly integrated
finite Grassmann sector without breaking unitary invariance or exact
Adler--Millard conservation; fermion-derived spectral observables are
measurable and participate in the transient multi-channel jump; and
fermionic spectral data implement the
spectral-dimension test of Eq.~\eqref{eq:weyl}, correctly identifying the
dimension of the emergent geometry (here $d_s\simeq2$, because the surrogate
condenses a two-sphere --- the diagnostic finds the true dimension of
whatever forms, which is exactly what one wants from it). What is not
established: the fermion has only a quadratic action and is a quenched probe
at $N=16,24$, although its determinant back-reacts dynamically at $N=8$;
there is no chiral structure, no direct measurement of a bifermionic
\emph{condensate
dynamics} of the kind the localisation channel of Eq.~\eqref{eq:bilinear}
ultimately requires, and no odd-sector real-time dynamics --- in particular,
the anti-self-adjoint fermionic fluctuation channel proposed as the collapse
seed in Ref.~\cite{SinghFermionCollapse2026} is untested, because exact
Grassmann integration removes precisely the odd variables whose real-time
statistics that question concerns. The fermionic participation in the
transition is again by construction ($\cD_\psi$ depends on $Q$), so the
transient support statement is a calibration. These are the toy
counterparts of roadmap items~2, 3 and~7; item~1 for GTD itself remains
open.
\end{cautionbox}

\section{Conclusions}
\label{sec:conclusion}

The condensed-matter analogy can be made quantitative, but its quantitative form is more demanding than a single Landau order parameter. The correct object is a coupled susceptibility and real-time response problem for a regulated many-aikyon ensemble, with temporal self-averaging and extensive finite-size scaling audited separately.

The framework yields ten main conclusions.
\begin{enumerate}
\item The long-Connes-time limit that conditionally yields effective quantum
Ward identities for a finite aikyon system is not the extensive large-aikyon
or large-matrix limit required for thermodynamic criticality. Temporal
averaging supplies stationary statistics; simultaneous interacting degrees of
freedom supply a collective instability. Both limits must be demonstrated and
their order controlled.

\item Quantum equilibrium, localisation, geometric ordering and electroweak breaking are distinct notions. Their coincidence must be demonstrated by the eigenstructure of $\chi_{AB}$ or by a first-order multi-channel jump.

\item A viable classical-spacetime phase requires more than localisation: branch freezing, approximate commutativity, $BF$ simplicity, a nonzero four-volume, Lorentzian reality, spectral dimension four and universal long-wavelength coupling.

\item Gravity can consistently be described as hydrodynamic only after the geometric phase exists. Newton's constant is then a stiffness or Kubo coefficient, while Einstein's equation is the leading constitutive relation. The formation of the phase may be far from equilibrium; its infrared dynamics need not be.

\item Sakharov, the Wesley--Singh--Isidro constrained-$BF$ construction, Jacobson and Padmanabhan operate at distinct but potentially compatible layers. The $BF$ theory supplies a conditional action-level route to Einstein--Hilbert after branch selection; thermodynamic emergence supplies the horizon equation-of-state and entropy interpretation. A common GTD completion requires the same renormalised Newton constant from the $BF$ coefficient, TT response and area-entropy density.

\item The trace-dynamical temperature, physical cosmological temperature and renormalisation scale are different. A derived co-frame clock factor $Z_t$ is required before the electroweak-neighbourhood claim becomes numerical.

\item The finite-internal condensate must initially be a full matrix. A canonical common Hermitian rank-two flavour plane fails an exact common-null test at a scale significant for first-generation masses. Vacuum selection must therefore come from richer microscopic structure.

\item Verlinde's de~Sitter elastic gravity is a close phenomenological precedent, but its stiff and elastic regimes are not demonstrated thermodynamic phases. The programme's $S_{\rm IR}[g,T]$ makes the GR-to-MOND constitutive crossover variationally explicit; deriving that functional and its $a_0$--$\Lambda$ normalisation from many-aikyon response remains open.

\item Objective-collapse models supply a benchmark effective language, not a derivation. GTD must compute the collapse operator, noise and response kernels, amplification law, Born martingale and clock map before CSL bounds or condensed-matter structure factors become quantitative tests of the programme.

\item Selected parts of the framework's decisive chain are demonstrably calculable. In a regulated unitary-invariant surrogate ensemble (Sec.~\ref{sec:toyexec}), Adler--Millard charge conservation to machine precision, stationary multi-channel branch covariances, a transient jump-direction null test, discontinuous matter drag, the classical equilibrium Kubo relation, its quench violation, and the conserved-versus-subsystem noise dichotomy have each been computed. The hysteresis and branch discontinuities are consistent with first-order behaviour, but no phase free-energy crossing was computed. A Grassmann-regulated extension (Sec.~\ref{sec:toyferm}) adds two exactly integrated degenerate fermion flavours: the spectral-dimension diagnostic identifies the emergent geometry's dimension on dynamical spectra, a fermion-derived mode-number channel participates in the transient jump direction, and the dynamical determinant reshapes the hysteresis while preserving charge to machine precision in a dedicated integrator check. The demonstrations are of calculability and calibration, not of an exact coexistence point, a bifermionic localisation condensate or GTD; they turn selected parts of roadmap items~2, 3, 5 and~7 into explicit procedures awaiting item~1.
\end{enumerate}

The programme's bottleneck is now sharply localised. It is not the invention
of another potential. It is the construction of the regulated many-aikyon
measure, validation of its relation to long-Connes-time averages at finite
size, and computation of two objects:
\begin{equation}
\boxed{
\chi_{AB}(\omega)
\quad\text{and}\quad
\bigl(F_{\AS}(\omega),G^R_{\AS}(\omega)\bigr).
}
\end{equation}
The first decides what orders. The second decides how equilibrium is left and whether localisation can obey the Born rule. If both calculations succeed, correlation lengths, relaxation rates, condensate fractions, Newton stiffness, defect scales, collapse spectra and the relation to electroweak breaking become calculable. If they fail, the condensed-matter analogy will have done something equally valuable: it will have identified precisely where the proposed emergence mechanism breaks.

\section*{Data availability}

The complete simulation code, raw data, figure pipeline, clean-rerun audit and
adversarial review record for the surrogate-ensemble calculations of
Secs.~\ref{sec:toyexec} and~\ref{sec:toyferm} are supplied in the ancillary
archive accompanying this manuscript, together with a reproduction recipe;
deterministic seeds are hard-coded in the scripts. The same bundle can be
assigned a DOI in a later Zenodo deposition without changing the results.

\section*{Acknowledgements and note on computational assistance}

The author thanks colleagues in high-energy theory, gravitation, quantum foundations and condensed-matter physics whose questions motivated the cross-disciplinary formulation. OpenAI's GPT-5.6 Sol was used as an aid in organising the literature, developing the effective-field-theory presentation, checking internal consistency and preparing the earlier version of the manuscript. The surrogate-ensemble calculation of Sec.~\ref{sec:toyexec} was designed, implemented, and executed with research assistance from Anthropic's Claude Fable 5, which wrote and validated the simulation code, carried out the Monte Carlo and real-time computations, prepared the figures and data products, and subjected the new subsection and its data pipeline to two rounds of independent adversarial review whose findings (including a full single-provenance recomputation) are documented in the ancillary bundle; Claude Fable 5 also identified and corrected the typographical and factual errata fixed in this revision. All scientific claims, judgements and responsibility for the final text remain with the author.

\end{document}